\documentclass[a4paper,10pt]{article}
\usepackage[utf8]{inputenc}

\usepackage{amssymb}
\usepackage{graphicx}
\usepackage{bm}
\usepackage{amsmath}
\usepackage{longtable}
\usepackage{lscape} 
\usepackage{color}
\usepackage[labelfont=bf,labelsep=period,justification=raggedright]{caption}

\title{Trophic coherence determines food-web stability}

\author{Samuel Johnson,$^{1\ast}$ Virginia Dom\'inguez-Garc\'ia,$^{2}$ Luca Donetti,$^{3}$ and\\Miguel A. Mu\~noz$^{2}$
\\
\small{$^{1}$Warwick Mathematics Institute, and Centre for Complexity Science,}\\
\small{University of Warwick, Coventry CV4 7AL, United Kingdom.}\\
\small{$^{3}$Departamento de Electromagnetismo y F\'isica de la Materia, and}\\
\small{Instituto Carlos I de F\'isica Te\'orica y Computacional,}\\
\small{Universidad de Granada, 18071 Granada, Spain.}\\
\small{$^{4}$Departamento de Electr\'onica y Tecnolog\'ia de Computadores, and CITIC-UGR,}\\
\small{Universidad de Granada, 18071 Granada, Spain.}\\
\small{$^\ast$E-mail:  S.Johnson.2@warwick.ac.uk}}

\date{}

\begin{document}

      \maketitle

\begin{abstract}

Why are large, complex ecosystems stable? 
Both theory and simulations of current models predict the onset of instability with growing size and complexity,
so for decades it has been
conjectured that ecosystems must have some unidentified structural property exempting them from this outcome.
We show that {\it trophic coherence} -- a hitherto ignored feature of food webs
which current structural models fail to reproduce -- 
is a better statistical predictor of
linear stability than size or complexity.
Furthermore, we prove that a maximally coherent network with constant interaction strengths will always be linearly stable.
We also propose a simple model which, by correctly capturing the
trophic coherence of food webs, accurately reproduces their
stability and other basic structural features.
Most remarkably, our model shows that stability can
increase with size and complexity. This suggests a key to May's
Paradox, and a range of opportunities and concerns for
biodiversity conservation.
\end{abstract}


\section*{Significance statement}

The fact that large, complex ecosystems are particularly robust is mysterious in the light of mathematical 
arguments which suggest they should be unstable – i.e. susceptible to runaway fluctuations in species' abundances. 
Here we show that food webs (networks describing who eats whom in an ecosystem) exhibit a property we call 
{\it trophic coherence}, a measure of how neatly the species fall into distinct levels. 
We find that this property makes networks far more linearly stable than if the links (predator-prey interactions) 
were placed randomly between species, or according to existing structural models. A simple model we propose 
to capture this feature shows that networks can, in fact, become more stable with size and complexity, 
suggesting a possible solution to the paradox.

\section*{Introduction}

In the early seventies, Robert May addressed the question of whether a generic system of coupled dynamical elements
randomly connected to each other would be stable. He found that the larger and more interconnected the system, 
the more difficult it would be to stabilise \cite{May,May_book}.
His deduction followed from the behaviour of the leading eigenvalue of the interaction matrix, which,
in a randomly wired system, grows with the square root of the mean number of links per element.
This result clashed with the received wisdom in ecology -- that large, complex ecosystems were 
particularly stable -- and initiated the ``diversity-stability debate'' \cite{MacArthur,Paine,Debate,Brose_book}.
Indeed, Charles Elton had expressed the prevailing view in 1958:
``the balance of relatively simple communities of plants and animals is more easily upset
than that of richer ones; that is, more subject to destructive
oscillations in populations, especially of animals, and more vulnerable to
invasions'' \cite{Elton_invasions}.
Even if this description were not accurate, the mere existence of rainforests and coral reefs
seems incongruous with a general mathematical principle that ``complexity begets instability'', and has become known 
as May's Paradox.

One solution
might be that the linear stability  
analysis used by May and many subsequent studies does not capture essential characteristics of ecosystem dynamics, and much work has 
gone into exploring how more accurate dynamical descriptions might enhance stability \cite{DeAngelis,Debate,neodinamic}.
But as ever better ecological data are gathered, it is 
becoming apparent that the leading eigenvalues of matrices related to food webs (networks in which the species are nodes and the links 
represent predation) 
do not exhibit the expected dependence on size or link density \cite{No_complexity}. Food webs must, therefore, have some 
unknown structural feature which accounts for this deviation from randomness -- irrespectively of other stabilising factors.

We show here that a network feature we call {\it trophic coherence} accounts for much of the variance in linear stability observed in a 
dataset of 46 food webs, and we prove that a perfectly coherent network with constant link strengths will always be stable.
Furthermore, a simple model that we propose to capture this property suggests that networks can become more 
stable with size and complexity if they are sufficiently coherent.

\section*{Results}

\subsection*{Trophic coherence and stability}
Each species in an ecosystem is generally influenced by others, via processes such as predation, parasitism, mutualism or 
competition for various resources \cite{Elton,Pimm,Jordi_book,100_Ecoquestions}. A food web is a network of species which 
represents the first kind of influence with directed links 
(arrows) from each prey node to its predators \cite{Pimm_82,Dunne,Drossel,Rossberg_book}.
Such representations can therefore be seen as transport networks, 
where biomass originates 
in the basal species (the sources) and flows through the ecosystem,
some of it reaching
the apex predators (the sinks).

The trophic level of a species can be defined as the average trophic level of its prey, 
plus one \cite{Levine_levels,Pauly_fishing_down}. 
Thus, plants and 
other basal species are assigned level one, pure herbivores have level two, but many species will have fractional 
values.\footnote{In computing the mean trophic level, it is customary to weight the contribution of each prey species by the 
fraction of the predator's diet that it makes up. Since we are here only considering binary networks, we do not perform this weighting.
We also use the words predator and prey as synonyms of consumer and resource, respectively, even in referring, say, 
to plants and herbivores.}
A species' trophic level provides a useful measure of how far it is from the sources of biomass in its ecosystem.
We can characterise each link in a network with a {\it trophic distance}, defined as the difference between the trophic 
levels of the 
predator and prey species involved (it is not a true ``distance'' in the mathematical sense, since it can be negative). 
We then look at the distribution of trophic distances over all links in a given network. The mean of 
this distribution will always be equal to one, while we refer to its degree of homogeneity as the network's {\it trophic coherence}.
We shall measure this degree of order with the standard deviation of the distribution of trophic distances, $q$ (we avoid using the 
symbol $\sigma$ since it is often assigned to the standard deviation in link strengths). A perfectly coherent 
network, in which all distances are equal to one (implying that each species occupies an integer trophic level),
has $q=0$, while less coherent networks 
have $q>0$. We therefore refer to this $q$ as an ``incoherence parameter''.
(For a technical description of these measures, see Methods.)

A fundamental property of ecosystems is their ability to endure over time \cite{Jordi_book,Rossberg_book}.
``Stability'' is often used as a generic term for any measure 
of this characteristic, including for concepts such as robustness and resilience \cite{Grimm_Babel}.
When the analysis regards the possibility that a small 
perturbation in population densities could amplify into runaway fluctuations, stability is usually understood in the sense of Lyapunov 
stability -- which in practice tends to mean linear stability \cite{Stability_scholarpedia}.
This is the sense we shall be interested in here, and henceforth ``stability'' will mean ``linear stability''.
Given the equations for the dynamics of the system, a fixed (or equilibrium) point will be linearly stable if all the 
eigenvalues of the Jacobian matrix evaluated at this point have negative real part.
Even without precise knowledge of the dynamics, one can still apply this reasoning to learn about the stability of a 
system just from the network structure of interactions between elements (in this case, species whose trophic interactions are described 
by a food web) \cite{May_book,May_qualitative,logofet_stability,logofet_book}.
In Methods (and, more extensively, in Section 3.2 of Supporting Information), we describe how an interaction matrix 
$W$ can be derived from the adjacency (or predation) matrix $A$ representing a food web, such that the real part of $W$'s leading 
eigenvalue, $R=Re(\lambda_1)$, is a measure of the degree of self-regulation each species would require in order for the system to 
be linearly stable. In other words, the larger $R$, the more unstable the food web.
For the simple yet ecologically unrealistic case in which the extent to which a predator consumes a prey species is proportional 
to the sum of their (biomass) densities, the Jacobian coincides with $W$, and $R$ describes the stability for any configuration of 
densities (global stability). For more realistic dynamics -- such as Lotka-Volterra, type II or type III -- 
the Jacobian must be 
evaluated at a given point, but we show that the general form can still be related to $W$ (see Methods).
Furthermore, by making assumptions about the biomass distribution, 
it is possible to check our results for such dynamics (see Section 3.2.1 of Supporting Information).
In the main text, however, we shall focus simply on the matrix $W$ without making 
any further assumptions about dynamics or biomass distributions.

May considered a generic Jacobian in which link strengths were drawn from a random distribution, representing all kinds of ecological interactions \cite{May,May_book}.
Because, in this setting, the expected value of 
the real part of the leading eigenvalue ($R$)
should grow with $\sqrt{SC}$, where $S$ is the number of species and $C$ the 
probability that a pair of them be connected, larger and more interconnected ecosystems should be less stable than small, 
sparse ones \cite{Sole00}.
(Allesina and Tang have recently 
obtained stability criteria for random networks with specific kinds of interactions: 
although predator-prey relationships 
are more conducive to stability than competition or mutualism, even a network consisting only of predator-prey interactions 
should become more unstable with increasing size and link density \cite{Allesina_Tang}.)

\begin{figure}[ht!]
\begin{center}
\includegraphics[scale=0.22]{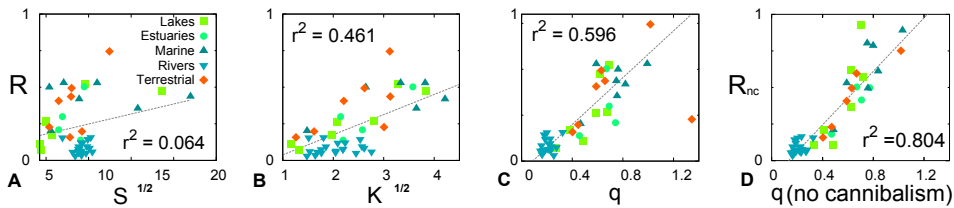}
\end{center}
\caption{
Scatter plots of stability (as measured by $R$, the real part of the leading eigenvalue of the interaction matrix) against
several network properties in a dataset of 46 food webs; Pearson's correlation coefficient is shown in each case.
{\bf A}: Stability against $\sqrt{S}$, where
$S$ is the number of species ($r^2=0.064$).
{\bf B} Stability against $\sqrt{K}$, where $K$ is the mean degree ($r^2=0.461$).
{\bf C} Stability against incoherence parameter $q$ ($r^2=0.596$).
{\bf B} Stability after all self-links (representing cannibalism) have been removed ($R_{nc}$) against incoherence parameter $q$
($r^2=0.804$).}
\label{fig_1}
\end{figure}

We analyse the stability for each of a set of 46 empirical food webs from several kinds 
of ecosystem (the details and references for these can be found in Section 2 of Supporting Information). 
In Fig. \ref{fig_1}A we plot the $R$ of each web against 
$\sqrt{S}$, observing no significant correlation. Figure \ref{fig_1}B shows $R$ against $\sqrt{K}$, where $K=SC$ is a network's 
{\it mean degree} (often referred to as ``complexity''). In contrast to a recent study by Jacquet {\it et al.} \cite{No_complexity}, 
who in their set of food webs found no significant complexity-stability relationship, we observe a positive correlation between 
$R$ and $\sqrt{K}$. 
However, less than half the variance in stability can be accounted for in this way. 
In Section 3.2.5 of Supporting Information we also compare the empirical $R$ values to the estimate derived by Allesina and Tang
for random networks in which all links are predator-prey. Surprisingly, the correlation is lower than for $\sqrt{K}$ ($r^2=0.230$).
The conclusion of Jacquet and colleagues -- 
namely, that food webs must have some non-trivial structural feature which explains 
their departure from predictions for random 
graphs -- therefore seems robust.

Might this feature be trophic coherence?
In Fig. \ref{fig_1}C we plot $R$ for the same food webs against the incoherence parameter $q$. 
The correlation is significantly stronger than with complexity -- stability increases with coherence.
However, there are still outliers, such as the food web of Coachella Valley.
We note that although most forms of intra-species competition
are not described by the interaction matrix, there is one form which is: cannibalism. This fairly common practice is a 
well-known kind of 
self-regulation which contributes to the stability of a food web (mathematically, negative elements in the diagonal of the interaction 
matrix shift its eigenvalues leftwards along the real axis). In Fig. \ref{fig_1}D we therefore plot the $R$ and $q$ we obtain after 
removing all self-links.
Now Pearson's correlation coefficient is $r^2=0.804$. In other words, cannibalism and trophic coherence together account 
for over $80\%$ of the variation in stability observed in this dataset.
In contrast, when we compare stability without self-links to the other measures, we find that for $\sqrt{S}$ the correlation becomes
negative (though insignificant), for $\sqrt{K}$ it rises very slightly to $r^2=0.508$, and for Allesina and Tang's estimate it drops 
below significance (see Section 3.2.5 of Supporting Information). In Section 3.2.1 of Supporting Information, we measure 
stability according to Lotka-Volterra, type II and type III dynamics, and show that in every case trophic coherence is the best predictor
of stability.

\subsection*{Modelling food-web structure}

Many mathematical models have been put forward to simulate various aspects of 
food webs \cite{Caldarelli,Sole00,Bastolla_models,Loeuille,Rossberg_experts,evolution1,Allesina_groups,Rossberg,Rossberg_book}.
We shall focus here on so-called structural, or static, models: those which attempt to reproduce 
properties of
food-web structure 
with a few simple rules. The best known is 
Williams and Martinez's Niche Model 
\cite{Williams_niche_model,nicheforever}. This is an elegant way of generating non-trivial networks by randomly assigning each species 
to a position on a ``niche axis'', 
together with a range of axis centred at some lower niche value. Each species then consumes all other species lying within its range
of axis, and none without. 
The idea is that the axis represents some intrinsic hierarchy among species which determines who can prey on whom.
The Niche Model is itself based on Cohen and Newman's Cascade Model,
which also has an axis, but
species are randomly assigned prey from amongst all those with lower niche values than themselves \cite{Cohen_1}.
Stouffer and colleagues proposed the 
Generalized Niche Model, in which some of a species' prey are set according to the Niche Model 
while the rest ensue from a slightly refined version of the Cascade Model,\footnote{The Generalized Cascade Model is like the original 
Cascade Model except that the numbers of prey species are drawn from the Beta distribution used in the Niche Model and subsequent 
niche-based models \cite{Stouffer_GNM}. This is the version of the model used throughout this paper, as explained in Section 1.1 of 
Supporting Information.}
the proportion of each being determined by a {\it contiguity} 
parameter \cite{Stouffer_robust}.
The Minimum Potential Niche Model of Allesina and co-workers is similar, but includes (random) forbidden links within 
species' ranges, instead of extra ones, as a way of emulating the effects of more than one axis
-- with the advantage that all the links of real food webs have a non-zero 
probability of being generated by this model \cite{Allesina}. 
Meanwhile, the Nested Hierarchy Model of Cattin {\it et al.}
takes into account that phylogenetically close species are more likely to share prey than unrelated ones \cite{Cattin}.
(For details of the models, see Section 1 of Supporting Information.)

These models produce networks with many of the statistical properties of food webs \cite{nicheforever,Stouffer_GNM,Allesina}. 
However, as we go on to show below, they tend to predict significantly less trophic coherence (larger $q$)
than we observe in our dataset. We therefore propose the Preferential Preying Model (PPM) as a way of capturing this feature.
We begin with $B$ nodes (basal species) and no links. We then add new nodes (consumer 
species) sequentially to the system until we have a total of $S$ species, assigning each their prey from amongst available nodes 
in the following way. The first prey species 
is chosen randomly, and the rest are chosen with a probability that decays exponentially with their absolute trophic distance to 
that initial prey species (i.e. with the absolute difference of trophic levels).
This probability is set by a parameter $T$ that determines the degree of trophic specialization of consumers.
The number 
of prey is drawn from a Beta distribution with a mean value proportional to the number of available species,
just as the other structural models described use a mean value proportional to the niche value.
(For a more detailed description, see Methods.)

The PPM is reminiscent of Barab\'asi and Albert's model of evolving networks \cite{Barabasi},
but it is also akin to a highly simplified version of an ``assembly model'' in which species enter via 
immigration \cite{Bastolla_models,evolution1}. 
It assumes that if a given species has adapted to prey 
off species A, it is more likely to be able to consume species B as well if A and B have similar trophic levels than if not.
It may seem that this scheme is similar in essence to the Niche Model, with the role of 
niche-axis being played by the trophic levels. However, 
whereas the niche values given to species in niche-based models are hidden variables, meant to represent some kind of biological 
magnitude, the trophic level of a node is defined by the emerging network architecture itself.
We shall see that this difference has a crucial effect on the networks generated by each model.

\begin{figure}[ht!]
\begin{center}
\includegraphics[scale=0.45]{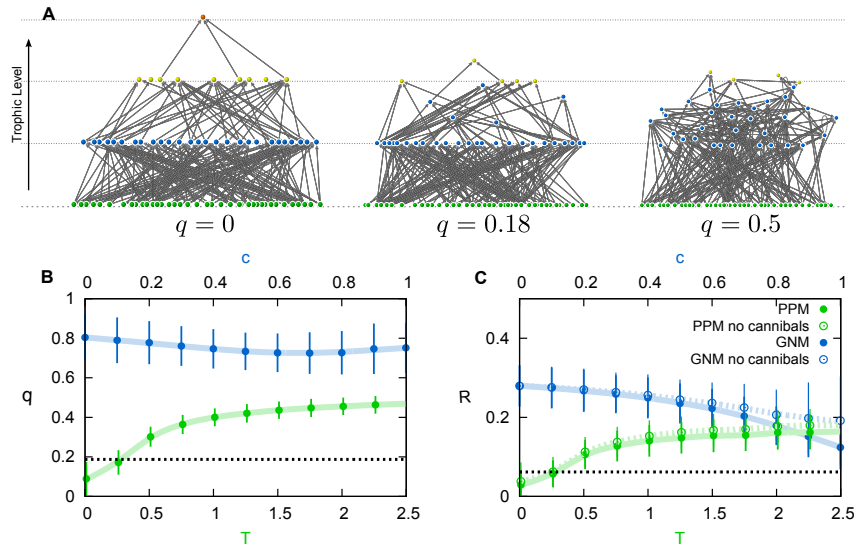}
\end{center}
\caption{
{\bf A}: Three networks with differing trophic coherence, the height of each node representing its trophic level. The networks 
on the left and right were generated with the Preferential Preying Model (PPM), with $T=0.01$ and $T=10$, respectively, yielding 
a maximally coherent structure ($q=0$) and a highly incoherent one ($q=0.5$). The network in the middle is the food web of a stream in 
Troy, Maine, which has $q=0.18$ \cite{Townsend_continent}. All three have the same numbers of species, basal species and links.
{\bf B}: Incoherence parameter, $q$, against $T$ for PPM networks with the parameters of the Troy food web (green); and against $c$ for 
Generalized Niche Model networks with the same parameters (blue). The dashed line indicates the empirical value of $q$.
{\bf C}: Stability (as given by $R$, the real part of the leading eigenvalue of the interaction matrix) for 
the networks of panel {\bf B}. Also shown is the stability of networks generated with the same models and parameters, but after removing 
self-links (empty circles). In panels {\bf B} and {\bf C}, the dashed line represents the empirical value of $R$, while bars on the 
symbols are for one standard deviation.
}
\label{fig_2}
\end{figure}

\subsection*{The origins of stability}

Figure \ref{fig_2}A shows three networks with varying degrees of trophic coherence. The one on the left was generated 
with the PPM and $T=0.01$, and since it falls into perfectly ordered, integer trophic levels, it is maximally coherent, with 
$q=0$. For the one on the right we have used $T=10$, yielding a highly incoherent structure, with $q=0.5$. Between these
two extremes we show the empirical food web of a stream in Troy, Maine \cite{Townsend_continent},
which has the same number of basal species, consumers and 
links as the two artificial networks, and an intermediate trophic coherence of $q=0.18$. 
Figure \ref{fig_2}B shows how trophic coherence varies with $T$ in PPM networks.
At about $T=0.25$ we obtain the empirical trophic coherence of the Troy food web (indicated with a dashed line). 
We also plot $q$ for networks generated with 
the Generalized Niche Model against ``diet contiguity'', $c$, its only free parameter \cite{Stouffer_robust}.
At $c=0$ and $c=1$ we recover the Cascade and Niche Models,
respectively (see Section 1.4 of Supporting Information).
However, diet contiguity has little effect on trophic coherence.

Figure \ref{fig_2}C shows the stability -- as measured by $R$, the leading eigenvalue of the interaction matrix -- for the networks 
of Fig. \ref{fig_2}B. For the PPM networks, stability closely mirrors trophic coherence: as $T$ decreases, the networks become 
more stable (smaller $R$) as well as more coherent (smaller $q$). The empirical value of $R$ is obtained 
at about the same $T$ which best approximates the empirical $q$. The Generalized Niche Model also generates more stable networks as
diet contiguity is increased, but this effect cannot be due to trophic coherence, which remains nearly constant.
The origin of increasing stability in this model is revealed when we measure $R_{nc}$
($R$ after removing all self-links from the networks): the Generalized Niche Model now 
displays only a very small dependence of stability on diet contiguity.
In contrast, the behaviour of $R_{nc}$ with $T$ in the PPM networks remains qualitatively 
the same as in the previous case, and the empirical stability continues to be obtained at $T\simeq 0.25$ 
(in this case, the empirical stabilities $R$ and $R_{nc}$ coincide, since the Troy food web has no cannibals).

\begin{figure}[ht!]
\begin{center}
\includegraphics[scale=0.32, angle=0]{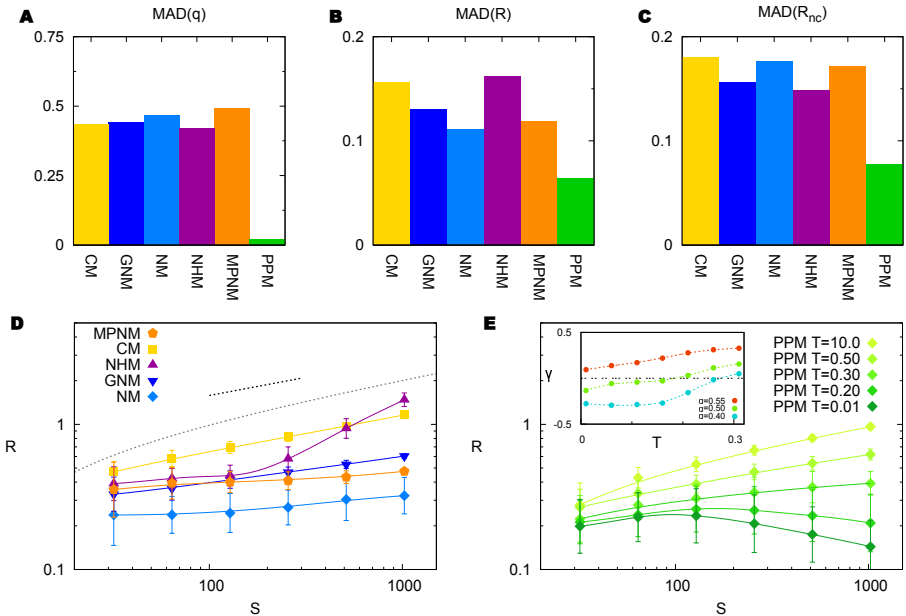}
\end{center}
\caption{
{\bf A}: Mean Absolute Deviations (MAD) from empirical values of the incoherence parameter, $q$, for each food-web model -- Cascade (CM), 
Generalized Niche (GNM), Niche (NM), Nested Hierarchy (NHM), Minimum Potential Niche (MPNM) and Preferential Preying (PPM) -- 
as compared to a dataset of 46 food webs.
{\bf B}: MAD from empirical values of stability, $R$, for the same models and food webs as in panel {\bf A}.
{\bf C}: MAD from empirical values of stability, $R$, after removing self-links, for the same models and food webs as in panels 
{\bf A} and {\bf B}.
{\bf D}: Scaling of stability, $R$, with size, $S$, in networks generated with each of the models of previous panels except for the PPM.
Mean degree is $K=\sqrt{S}$. 
The dashed line indicates the slope predicted for random matrices by May \cite{May}, while the dotted curve is from Allesina and Tang 
\cite{Allesina_Tang}.
{\bf E}: Scaling of stability, $R$, with size, $S$, in PPM networks generated with different values of $T$. 
In descending order, $T=10$, $0.5$, $0.3$, $0.2$ and $0.01$.
$B=0.25S$.
Inset: Slope, $\gamma$, of 
the stability-size line against $T$ for $\alpha=0.55$, $0.5$ and $0.4$, where the mean degree is $K=S^\alpha$.
In panels {\bf D} and {\bf E}, bars on the 
symbols are for one standard deviation.
}
\label{fig_3}
\end{figure}

We perform this analysis for each of the $46$ food webs in our dataset, obtaining the value of $T$ which best captures the empirical
trophic coherence according to the PPM. We then compute the ensemble averages of $R$ and $R_{nc}$ generated at this $T$, for comparison 
with the empirical values.
Similarly, we compute the average values of these measures predicted by each of the niche-based models described above -- the Cascade, 
Niche, Nested Hierarchy, Generalized Niche and Minimum Potential Niche Models. The last two models have free parameters, but as these
do not have a significant effect on trophic coherence, we use the values published as optimal in Refs. \cite{Stouffer_GNM} and 
\cite{Allesina}, respectively (or the mean optimal values for those food webs which were not analysed in these papers).
Figures \ref{fig_3}A-C show the average absolute deviations from the empirical values for trophic coherence and stability, before 
and after removing self-links, for each model. In Fig. \ref{fig_3}A we observe that, as mentioned above, the niche-based models fail 
to capture the trophic coherence of these food webs. Stability, whether with or without considering self-links, is
predicted by the PPM significantly better than by any of the other models, as shown in Fig. \ref{fig_3}B and Fig. \ref{fig_3}C. 
This is in keeping with Alessina and Tang's observation that current structural models cannot account for food-web 
stability \cite{Allesina_Tang}.
In Section 3 of Supporting Information we show the results of similar model comparisons for several other network measures: 
modularity, mean chain length, mean trophic level, and numbers of cannibals and of apex predators. The PPM does as well as any of the
other models as regards numbers of cannibals and apex predators, and is significantly better at predicting the other 
measures.\footnote{Allesina and co-workers have developed a likelihood-based approach for comparing food-web models \cite{Allesina}.
We have not yet been able to obtain the corresponding likelihoods for the PPM, but if this is done in the future it would provide a firmer basis from which 
to gauge the models' relative merits, and perhaps to build a more realistic model drawing on each one's strengths.}

Why does the trophic coherence of networks determine their stability? 
The case of a maximally coherent structure, with $q=0$ (such as the one on the left in Fig. \ref{fig_2}A), 
is amenable to mathematical analysis. 
In Section 4 of Supporting Information we consider the undirected network that results from replacing each directed link of the 
predation matrix with a 
symmetric link, the non-zero eigenvalues of which always come in pairs of real numbers $\pm \mu_j$.
We use this to prove that the eigenvalues of the interaction matrix we are actually interested in, if $q=0$, will in turn come 
in pairs $\lambda_j=\pm \sqrt{-\eta} \mu_j$, where $\eta$ is a parameter related to the efficiency of predation
(considered, for the proof, constant for all pairs of species).
All the eigenvalues will therefore be real if $\eta<0$, zero if $\eta=0$,
and imaginary if $\eta>0$. A positive $\eta$ is the situation which corresponds to a food web -- or any system in which 
the gain in a ``predator'' is accompanied by some degree of loss in its ``prey''. Therefore, a perfectly coherent network is a 
limiting case 
which can be stabilised by an infinitesimal degree of 
self-regulation (such as cannibalism or other intra-species competition).
Any realistic situation would involve some 
degree of self-regulation, so we can conclude that a maximally coherent food web with constant link strengths would be stable.

Although a general, analytical 
relationship between trophic coherence and stability remains elusive, it is intuitive to expect that a deviation from maximal coherence 
will drive the real part of the leading eigenvalue towards the positive values established for random structures, as is indeed 
observed in our simulations.

\subsection*{May's Paradox}

As we have seen, the PPM can predict the stability of a food web quite accurately just with information regarding numbers of species, 
basal species and links, and trophic coherence. 
But what does this tell us about May's Paradox -- the fact that large, complex ecosystems 
seem to be particularly stable despite theoretical predictions to the contrary?
To ascertain how stability scales with size, $S$, and complexity, $K$, in networks generated by different models, we must first 
determine how $K$ scales with $S$ -- i.e. if $K\sim S^\alpha$, what value should we use for $\alpha$?
Data in the real world are noisy in this regard, and both the ``link-species law'' ($\alpha=0$) 
and the ``constant connectance hypothesis'' ($\alpha=1$) have been defended in the past, although the most common
view seems to be that $\alpha$ lies somewhere between zero and $1/2$ \cite{Pimm,Sole00,Rossberg_universal}. 
The most recent empirical estimate we are 
aware of is close to $\alpha\simeq 0.5$, depending slightly on whether predation weights are considered \cite{connectance}.
In our dataset, the best fit is achieved with a slightly lower exponent, $\alpha=0.41$.

In Fig. \ref{fig_3}D we show how stability scales with $S$ in each of the niche-based models when complexity increases with size 
according to $\alpha=0.5$. The dashed line shows the slope that May predicted for random networks 
($R\sim \sqrt{K}= S^{0.25}$)
\cite{May}. 
We also plot the curve recently shown by Allesina and Tang to correspond to random networks in which all interactions are 
predator-prey \cite{Allesina_Tang}, which has a similar slope to May's at large $S$.
This scaling is 
indeed closely matched by the Cascade Model.
The behaviour of the other models is similar (except for the Nested Hierarchy Model, in 
which $R$ increases more rapidly at high $S$), and, as expected, networks always become 
less stable with increasing size and complexity. In Fig. \ref{fig_3}E we show how the stability of PPM networks scales in the same 
scenario. For high $T$, their behaviour is similar to that of the Cascade Model: $R\sim S^{\gamma}$, 
with $\gamma\simeq 0.25$. However, the exponent $\gamma$ decreases as $T$ is lowered, until, for sufficiently 
large and coherent networks, 
it becomes negative -- in other words, {\it stability increases with size and complexity}.
The inset in Fig. \ref{fig_3}E shows the exponent $\gamma$ obtained against $T$, for different values of $\alpha$. The smaller 
$\alpha$, the larger the range of $T$ which yields a positive complexity-stability relationship.\footnote{Plitzko and 
colleagues recently showed that there exists a range of parameters (in a Generalized Modeling framework \cite{Gross}) 
for which Niche Model networks can increase in stability with complexity \cite{Drossel_complexity}. However, for this study 
networks were rejected unless they were stable and had exactly four trophic levels. This selection may have screened for 
trophic coherence, cannibalism or other structural features.}

In Section 3.2.1 of Supporting Information we extend this analysis to specific dynamics -- Lotka-Volterra, type II and type III --
by assuming an exponential relationship between biomass and trophic level which can be described as a pyramid.
The positive complexity-stability relationship does 
not appear to depend on the details of dynamics. However, the slope of the $R-S$ curve varies with both the squatness of the 
biomass pyramid and the extent to which the pyramid is corrupted by noise. A squat pyramid (more biomass at low trophic levels than 
at high ones) has the strongest relationship, while 
for an inverted pyramid (more biomass at high trophic levels than at low ones) the slope can flatten out or change sign.
Noise in the biomass pyramid tends always to weaken the positive complexity-stability relationship, and can also change its sign.

\section*{Discussion}

The predation matrices corresponding to real ecosystems are clearly peculiar in some way, since their largest eigenvalues do not
depend solely on their size or complexity, as we would expect both from random graph theory and structural food-web models. This 
is in keeping with the empirical observation that large, complex ecosystems are particularly stable, but challenges current thinking 
on food-web architecture. We have shown that the structural property we call trophic coherence is significantly correlated with food-web 
stability, despite other differences between the ecosystems and the variety of empirical methods used in gathering the data. 
In fact, cannibalism and trophic coherence together account for most of the variance in stability observed in our dataset.
Furthermore, we have proved that a maximally coherent food web with constant interaction strengths will always be stable.

We have suggested the Preferential Preying Model as a simple algorithm for generating networks with tunable trophic coherence. 
Although this model does not attempt to replicate other characteristic features of food webs -- such as a phylogenetic signal 
or body-size effects -- it reproduces the empirical stability of the 46 webs analysed quite accurately once its only 
free parameter has been adjusted to the empirical degree of trophic coherence. Most remarkably, the model predicts that 
networks should become more stable with increasing size and complexity, as long as they are sufficiently coherent and the number of 
links
does not grow too fast with size. Although this result should be followed up with further analytical and empirical research, it 
suggests that we need no longer be surprised at the high stability of large, complex ecosystems.

We must caution that these findings do not imply that trophic coherence 
was somehow selected for by the forces of nature in 
order to improve food-web function.
It seems unlikely that there should be any selective pressure on the individuals making up a species to do what is best for 
their ecosystem. 
Rather, many biological features of a species 
are associated with its trophic level. Therefore, adaptations 
which allow a given predator to prey on species A are likely to be useful also in preying on species B if A and B 
have similar trophic levels. This leads to trophic coherence, which results in high stability.

If stability decreased with size and complexity, as previous theoretical studies have assumed, ecosystems could not grow 
indefinitely, for they would face a cut-off point beyond which they would become unstable \cite{Sole00}.
On the other hand,
if real ecosystems are coherent enough that they become more stable with size and complexity, as our model predicts, 
then the reverse might be true.
We must also bear in mind, however, that our results are only for linear stability, whereas {\it structural} stability,
for instance, may depend differently on size and coherence, and could become the limiting factor \cite{Rossberg_book}.
In any case, ascertaining whether the loss of a few species would stabilise or destabilise a given community could be important 
for conservation efforts, particularly for averting ``tipping points'' \cite{100_Ecoquestions}.

The findings we report here came about by studying food webs. However, directed networks of many kinds transport energy, matter, 
information, capital or other entities in a similar way to how food webs carry biomass from producers to apex predators.
It seems likely that the relation between a network's trophic coherence and its leading eigenvalue will be of consequence to other 
disciplines, and perhaps 
the Preferential Preying Model, though overly simplistic for many scenarios, may serve as a first approximation for looking into 
these effects in a variety of systems.

\section*{Methods}

\subsection*{Measuring stability}

Let us assume that the 
populations of species making up an ecosystem 
(each characterised by its total biomass) 
change through time according to some set of nonlinear differential equations, the 
interactions determined by the predation matrix, $A$ (whose elements $a_{ij}$ take the value one if species $i$ preys on species $j$, 
and zero otherwise). If the system 
persists
without suffering large changes 
it must, one assumes, find itself in the 
neighbourhood of a fixed point of the dynamics. We can study how the system would react to a small perturbation by 
expanding the equations of motion around this fixed point and keeping only linear terms. The subsequent effect of the perturbation 
is then determined by the corresponding Jacobian matrix, and the system will tend to return to the fixed point only if the real parts of all its 
eigenvalues are negative \cite{Stability_scholarpedia}.

Even without knowledge of the details of the dynamics, it is 
possible to 
draw some conclusions
about the stability of a food web solely from its predation matrix \cite{May_qualitative}. 
Independently of the exact interaction strengths, we know that not all the biomass lost by a prey species when consumed 
goes to form part of the predator -- in fact, this efficiency is relatively low \cite{efficiency}. 
It is therefore natural to assume that 
the effect of species $j$ on species $i$ will be mediated by $w_{ij}=\eta a_{ij} - a_{ji}$, where $\eta$ is an efficiency 
parameter which, without further information, we can consider equal for all pairs of species. We can thus treat the 
interaction matrix $W=\eta A - A^T$ as the Jacobian of some unspecified dynamics. However, we have ignored the stabilising
effect of intra-species competition -- the fact that individuals within a species compete with each other
in ways 
which are not specified by the predation matrix.
This would correspond to real values to be subtracted from the
diagonal elements of $W$, thereby shifting its 
set of eigenvalues (or spectrum)
leftwards along the real axis. Therefore,
the eigenvalue with largest real 
part of $W$, as defined above, can be seen as a measure of the minimum intra-species competition required for the system to be stable. 
Thus, the lower this value, $R=Re(\lambda_1)$, the higher the stability.

In Section 3.2 of Supporting Information, we describe this analysis in more detail. Beginning with a general consumer-resource 
differential equation for the biomass of each species, we obtain the Jacobian in terms of the function $F(x_i,x_j)$ which describes
the extent to which species $i$ consumes species $j$. For the simple (and unrealistic) case $F=x_i+x_j$, the Jacobian reduces 
to the matrix $W$ as given above, independently of the fixed point. For more realistic dynamics, the Jacobian depends on the fixed point.
For instance, for the Lotka-Volterra function $F=x_i x_j$, the off-diagonal elements of the Jacobian are $J_{ij}=w_{ij}x_i$.
If we set $F=x_iH(x_j)$ (with $H(x)=x^h/(x^h+x_0^h)$, $x_0$ the half-saturation density and $h$ the Hill coefficient), we 
have either type II ($h=1$) or type III ($h=2$) dynamics \cite{Real}. Then the off-diagonal elements are 
$J_{ij}=[\tilde{\eta}(x_i,x_j)a_{ij}-a_{ji}]H(x_i)$, where the effective efficiency is 
$\tilde{\eta}(x_i,x_j)= \eta h x_0^h x_i x_j^{-(h+1)} H(x_j)^2 / H(x_i) $.

The Jacobians for Lotka-Volterra, type II and type III dynamics are all similar in form to the matrix $W$, although for an exact solution 
we require the fixed point. In the main text we therefore use the leading eigenvalue of $W$ as a generic measure of stability. However,
in Section 3.2.1 of Supporting Information we consider the effects that different kinds of biomass distribution have on each of 
these more realistic 
dynamics. The results are qualitatively the same as those for the matrix $W$, although we find that both the squatness of a biomass 
pyramid and the level of noise in this structure affect the strength of the diversity-stability relationship described in the main text.

This measure of stability depends on the parameter $\eta$. 
In Section 3.2.2 of Supporting Information we show that the results reported here remain qualitatively unchanged for 
any $\eta\in(0,1)$, and discuss 
how stability is affected when we consider $\eta>1$ or $\eta<0$. We also look into the effects of including a noise term so 
that $\eta$ does not have the same value for each pair of species, and find that our results are robust to this change too.
For the results in the main text, however, we use the fixed value $\eta=0.2$.

\subsection*{Trophic levels and coherence}

The trophic level $s_i$ of species $i$ is defined as the average trophic level of its 
prey, plus one \cite{Levine_levels}. That is, 
\begin{equation}
s_{i} = 1+\frac{1}{k_{i}^{in}}\sum_j a_{ij}s_{j},
\label{eq_si}
\end{equation}
where $k_{i}^{in}=\sum_j a_{ij}$ is the number of prey
of species $i$ (or $i$'s {\it in-degree}),
and $a_{ij}$ are elements of the predation matrix $A$. Basal species (those with $k^{in}=0$) are assigned $s=1$.
The trophic level of each species is therefore a purely structural (i.e. topological) property which can be determined by solving 
a system of linear equations. 
Since we only consider unweighted networks here (the elements of $A$ are ones and zeros), 
we omit the link strength term usually included in Eq. (\ref{eq_si}) \cite{Levine_levels}.

We can write Eq. (\ref{eq_si}) in terms of a modified graph Laplacian matrix,
$\Lambda {\bf s}={\bf v}$,
where ${\bf s}$ is the vector of trophic levels,
${\bf v}$ is the vector with elements $v_i=\mbox{max}(k_i^{in},1)$,
and $\Lambda=\mbox{diag}({\bf v})-A$.
Thus, every species can be assigned a trophic level if and only if $\Lambda$ is invertible. This requires at least one basal 
species (else zero would be an eigenvalue of $\Lambda$).
However, note that cycles are not, in general, a problem, despite the apparent recursivity of Eq. (\ref{eq_si})

We define the ``trophic distance'' spanned by each link $(a_{ij}=1)$ as $x_{ij}=s_i-s_j$
(which is not a distance in the mathematical sense since it can take negative values).
The distribution of trophic distances 
over the network is $p(x)$, which will have mean $\langle x\rangle=1$ (since for any node $i$ the average 
over its incoming links is $\sum_j a_{ij}(s_i-s_j)/k_i^{in}=1$ by definition). 
We define the ``trophic coherence'' of the network as the homogeneity of $p(x)$: the more similar the trophic distances 
of all the links, the more coherent. As a measure of coherence, we therefore use the 
standard deviation of the distribution, which we refer to as an incoherence parameter: $q=\sqrt{\langle x^2\rangle-1}$,
where $\langle\cdot \rangle=L^{-1}\sum_{ij}(\cdot)a_{ij}$, and $L$ is the total number of links, $L=\sum_{ij}a_{ij}$.

Trophic coherence bears a close resemblance to Levine's measures of ``trophic specialization'' \cite{Levine_levels}.
However, our average is computed over links instead of species, with the consequence that we need not consider the distinction
between resource and consumer specializations. It is also related to measures of omnivory: in general, the more omnivores one 
finds in a community, the less coherent the food web.

\subsection*{The Preferential Preying Model}

We begin with $B$ nodes (basal species) and no links. We then add, sequentially, $S-B$ new nodes 
(consumer species) to the system according to the following rule. A new node $i$ is first awarded a random 
node $j$ from among all those available when it arrives. Then another $\kappa_i$ nodes $l$ are chosen with 
a probability $P_{il}$ that decays with the trophic distance between $j$ and $l$. Specifically, we use 
the exponential form
$$
P_{il}\propto \exp\left(-\frac{|s_j-s_l|}{T}\right),
$$ 
where $j$ is the first node chosen by $i$, and $T$ is a parameter that sets the degree of trophic specialization 
of consumers.

The number of extra prey, $\kappa_i$, is obtained in a similar manner to the Niche Model prescription, since this has been shown 
to provide the best approximation to the in-degree distributions of food webs \cite{Stouffer_GNM}. We set $\kappa_i=x_i n_i$,
where $n_i$ is the number of nodes already in the network when $i$ arrives, and $x_i$ is a random variable drawn from a Beta distribution
with parameters
$$
\beta=\frac{S^2-B^2}{2L}-1,
$$
where $L$ is the expected number of links. In this work, we only consider networks with a number of links within an error margin of
$5\%$ of the desired $L$; thus, for all the results reported, we have imposed this filter on the PPM networks and those generated with 
the other models.

To allow for cannibalism, the new node $i$ is initially considered to have a 
trophic level $s_i=s_j+1$ according to which it might then choose itself as prey. Once $i$ has
been assigned all its prey, $s_i$ is updated to its correct value.

\section*{Acknowledgements}
We are grateful to J. Dunne and U. Jacob for providing food-web data and to J. Hidalgo for technical help. 
Many thanks also to N.S. Jones, D.C. Reuman, D.B. Stouffer, J. Bascompte, I. Mendoza, J.A. Bonachela, 
S.A. Levin, O. Al Hammal, A. Thierry, A. Maritan, N. Guisoni, G. Baglietto, 
E. Albano, B. Moglia, J.J. Torres, and J.A. Johnson for conversations and comments 
on versions of the manuscript. 
We acknowledge financial support from the Spanish MICINN-FEDER under project FIS2009-08451 
and from Junta de Andaluc{\'\i}a Proyecto de Excelencia P09FQM-4682. 
S.J. is grateful for support from the Oxford Centre for Integrative Systems Biology, 
University of Oxford; and from the European Commission under the Marie Curie Intra-European 
Fellowship Programme PIEF-GA-2010-276454.

\newpage



\end{document}


\maketitle

\newpage

\tableofcontents

\newpage

\setcounter{section}{0}

\section{Food-web models}
\label{Section_Models}

We describe here the main structural (also called {\it static}) models found in the literature for generating networks with some of 
the statistical features of food webs. We then discuss some aspects of the Preferential Preying Model (PPM) which we put forward in the 
main text (described in Methods). In all these models, the number of links $L$ can only be set in expected value. As is often done, 
throughout this work we discard all generated networks which have a number of links greater or smaller than this target $L$ by 
more than five percent. In Section \ref{Section_Measures} we describe several network measures and compare the performance of the 
models using the food-web data listed in Section \ref{Section_Data}.

\subsection{The Cascade Model}

In the Cascade Model, each species $i$ is assigned a random number $n_i$ drawn from a uniform distribution 
between $0$ and $1$ \cite{Cohen_1}. For any pair ($i$, $j$), we set $i$ to be a consumer of $j$ with a constant probability
$p$ if $n_i>n_j$, and with probability zero if $n_i\leq n_j$. With $S$ species, we obtain an expected number of links $L$
if we set
$$
p=\frac{2L}{S(S-1)}.
$$
This was the first attempt to show how networks with a structure in some senses similar to real food webs could come 
about via simple rules.

Stouffer and co-workers later modified this model so that the number of prey would be drawn from the Beta distribution 
used by the Niche Model 
(see below), and called the new version the Generalized Cascade Model \cite{Stouffer_GNM}. Since this amendment improves the 
model's predictions as regards distributions of prey and predators (without, to the best of our knowledge, involving any drawbacks),
throughout this paper we use the Generalized Cascade Model.

\subsection{The Niche Model}

In the Niche Model, each species $i$ is awarded a niche value $n_i$ as in the Cascade Model \cite{Williams_niche_model}. 
However, instead of choosing
species with lower niche values randomly for prey, $i$ is constrained to consume the subset of species $j$ such that
$c_i-r_i/2 \leq n_j < c_i+r_i/2$ -- i.e., all those lying on an interval of the niche axis of size $r_i$ and centred at $c_i$, 
and none without.
The range is defined as $r_i=x_i n_i$, where $x_i$ is drawn from a Beta distribution with parameters $(1,\beta)$. For $S$
species and a desired number of links $L$, we must set 
$$
\beta=\frac{S(S-1)}{2L}-1.
$$
The centre of the interval $c_i$ is drawn from a uniform distribution between $r_i/2$ and $min(n_i,1-r_i/2)$.

The rationale behind this model was that food webs were thought to be {\it interval} -- i.e., the species could be arranged 
in an ordering such that the prey of any given predator were contiguous \cite{Cohen_book}. The Niche Model achieves this by construction.
More recent analysis has shown that food webs are not generally perfectly interval, although they do usually exhibit a certain degree of 
intervality \cite{Stouffer_robust,Capitan}. Nevertheless, the Niche Model has been tremendously successful, since it outperforms the 
Cascade Model in approximating
measurable features of food webs, and even compares well to more elaborate models which take the Niche Model as a 
basis \cite{nicheforever}. It is still the model most commonly used whenever synthetic networks similar to food webs are required.

\subsection{The Nested Hierarchy Model}

The Nested Hierarchy Model provides a way to take into account that phylogenetically similar species should have prey in 
common \cite{Cattin}.
It gives each species a niche value and a range, exactly as in the Niche Model. However, instead of establishing links directly to 
species within the range, first the number of prey to be consumed by each species is determined, in proportion to the range, 
$k_i^{in}\propto r_i$, so as to generate an expected number of links $L$.
These links are then attributed in the following way.
The species with lowest niche value has no prey, while the one with the highest has no predators (so there is always at least 
one basal species and one apex predator). Starting from the species with second smallest niche value and going up in order
of $n$, we take each species $i$ and apply the following rules to determine its $k_i^{in}$ prey:
\\
1. We choose a random species $j$ already in the network (so $n_j\leq n_i$) and set it as the first prey species of $i$.
\\
2. If $j$ has no predators other than $i$, we repeat 1 until either the chosen prey does have other predators, or we 
reach $k_i^{in}$. Else we go to 3.
\\
3. We determine the set of species which are prey to the predators of $j$. We select, randomly, species from this set to 
become also prey of $i$ until we either complete $k_i^{in}$, or we go to 4.
\\
4. We continue choosing prey species randomly from among those with lower niche values. If we still have not reached $k_i^{in}$ when 
these run out, we continue choosing them randomly from those with higher niche values.

In this model, two consumers that share prey are assumed to be phylogenetically related, while the extra links that must at times be 
sought mimic the effects of independent adaptation. We find it a particularly interesting model because phylogenetic constraints should indeed be taken into 
account, and as it stands our Preferential Preying Model (described below) does not do this.
One problem we find with the Nested Hierarchy Model, however, is that a given species $i$ is assumed to be related to a certain set 
$A$ of species which share common prey with $i$; but $i$ will also belong to the set $B$ of common prey of a different set of consumers,
and nothing constrains $A$ and $B$ to overlap. In other words, the species related to $i$ due to its prey are not the 
ones related to $i$ due to its predators, whereas in nature it is to be expected that phylogenetically similar species should have 
both prey and predators in common. In fact, it has recently been reported that common predators are statistically more significant 
than common prey \cite{Rossberg_phylogeny}.



\subsection{The Generalized Niche Model}

The Generalized Niche Model was proposed to account for the fact that empirical food webs turned out not to be maximally
interval, as predicted by the Niche Model \cite{Stouffer_robust}. 
A {\it contiguity} parameter $c$ was introduced, which would determine the proportion 
of prey to be allocated according to the Niche Model, the rest ensuing from the Generalized Cascade Model. 
In other words, the Niche Model 
would be implemented as before but with reduced ranges $r_i= c x_i n_i$. Then, for each species, the number of extra prey 
$k_i^{cascade}=(1-c)x_i n_i S$ is drawn randomly from among the available species with niche values lower than $n_i$, as in the 
Generalized Cascade Model. For $c=1$ we have the Niche Model, while $c=0$ results in the Generalized Cascade Model. 

The Generalized Niche Model has been shown to emulate real food webs very successfully, at least as regards certain features, 
such as community structure \cite{Stouffer_community}. It is also often used as a convenient model for generating
synthetic networks with a view to studying food webs {\it in silico} \cite{Gross}.

\subsection{The Minimum Potential Niche Model}

The Minimum Potential Niche Model is similar to the Generalized Niche Model in that it is a modification of the Niche Model 
which breaks up complete intervality by means of a parameter, $f$ \cite{Allesina}. 
However, the motivation is slightly different. The idea is that in 
reality there is more than one niche dimension constraining possible predation links (hence the lack of complete, one-dimensional 
intervality), 
which implies that some of the links determined by the Niche Model are actually ``forbidden links''. The species are all allocated 
niche values $n_i$ and ranges $r_i=x_i n_i$ as in the Niche Model. The species at the extremes of this range are always consumed. 
However, the rest is considered a potential range and the $\beta$ 
parameter used in the Beta distribution from which $x_i$ is drawn is now
$$
\beta=\frac{S(S-1)}{2(L+F)}-1,
$$
where $F=fP$, $P$ being the total number of potential links given the ranges, minus the species at the extremes. Once all the species 
have their ranges, each species within will be consumed with a probability $1-f$. Therefore, $f=0$ results in the original Niche Model, 
but $f>0$ produces a proportion of forbidden links.

Allesina {\it et al.} suggested a framework for comparing niche-based models \cite{Allesina}; they computed the likelihood 
that the Cascade, Niche and Nested Hierarchy models have 
of generating the links in a set of ten real food webs, and found theirs (the Minimum Potential Niche Model) to be superior --
and, in fact, the only one capable of generating all the observed links.

\subsection{The Preferential Preying Model}

In the main text we propose the Preferential Preying Model (PPM) in order to capture the {\it trophic coherence} of empirical 
food webs. The details are given in Methods, so here we confine ourselves to displaying the scheme diagrammatically 
in Fig. S\ref{fig_model}. We go on to list several possible amendments which could be made to this basic version of the model and 
which may be of use to researchers wishing to use the PPM for purposes other than our main one here -- namely, to highlight the 
importance of trophic coherence and its relevance to food-web stability.

\begin{figure}[ht!]
\setcounter{figure}{0}
\renewcommand{\figurename}{Figure S}
\begin{center}
\includegraphics[scale=0.25]{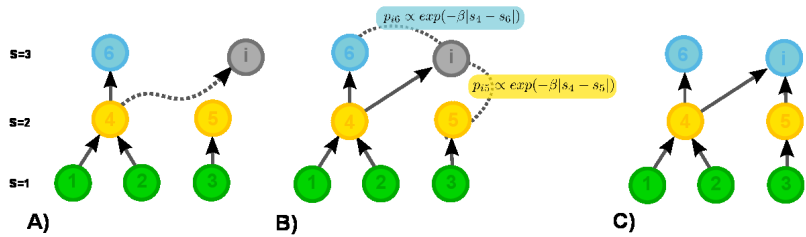}
\end{center}
\caption{
Diagram showing how networks are assembled in the Preferential Preying Model (PPM), as described in Methods in the main text. 
In Panel {\bf A} a new node, labelled $i$, is introduced to the networks, and is randomly assigned node $4$
as its first prey species. In Panel {\bf B}, the probabilities of next choosing node $5$ or node $6$ are calculated, as functions 
of their trophic distance to node $4$
($\beta=1/T$). Node $5$ is the closest, and in this case is taken as the second prey species, as shown in 
Panel {\bf C}. 
}
\label{fig_model}
\end{figure}


\subsubsection{Possible amendments to the PPM}


\begin{itemize}
 \item 
 {\bf Basal species.} All the niche-based models discussed allow the number of producers, $B$, to emerge freely
 (although they are not, generally, particularly successful in predicting $B$ \cite{nicheforever}).
 We chose here 
 to begin with a set number of basal species, as in the Preferential Attachment Model \cite{Barabasi}. We imagine that for most
 applications where synthetic networks are required it would be useful to have control over this parameter (which is itself related 
 to trophic coherence, as we show in Section \ref{sec_herbivory}).
 However, if a freely 
 emerging $B$ were preferred -- for instance, for a rigorous comparison against models which do not allow this value to be set easily --
 it is straightforward to take the minimum $\kappa_i$ equal to zero for incoming species, thereby allowing a proportion of them 
 to become producers.
 
 \item
 {\bf Numbers of prey.} We have drawn the number of prey for each incoming species from a Beta distribution, as in all the niche-based 
 models, because
 Stouffer {\it et al.} \cite{Stouffer_GNM} have shown that this method yields a particularly good fit to food-web data 
 (we have also verified that this holds true for our 46 food-web dataset). However, were the model to be applied to systems other than 
 food webs, it may be preferable to use, for instance, a Poisson or a Pareto distribution, depending on the in-degree distributions of 
 the networks to be emulated.

 \item
 {\bf Boltzmann factor.} The functional form we have used to determine the second and subsequent prey of an incoming species  
 (an exponential in the trophic distance divided by the parameter $T$) is arbitrary; careful fitting to data may suggest a better 
 function. There is also no reason other than simplicity to use the same value of $T$ for each incoming species: one could also 
 draw a different value $T_i$ for each incoming species form some distribution, perhaps dependent on the trophic level of its first prey.
 
 \item
 {\bf Cycles}. Directed loops in food webs are relatively rare, yet often present. The PPM as described does not generate any of these 
 cycles, but it could easily be amended to do so by assigning each incoming species a small number of predators as well as prey 
 from amongst the species already in the network. However, directed loops require some predators to consume prey at higher trophic levels 
 than theirs, so the more coherent a network, the fewer directed loops are to be expected.
 
 \item
 {\bf Phylogeny and body size.} In this simple incarnation, the PPM ignores the main effects that most of the other models are based 
 on, but these could be taken into account in a ``Generalized Preferential Preying Model''. Something akin to a
 phylogenetic signal
 could be induced by introducing a bias in the Boltzmann factor such that an incoming node tended to copy the prey and predators of 
 a randomly chosen species already in the network -- perhaps limiting in the Nested Hierarchy Model in the case where only prey are 
 copied. The Niche, Generalized Niche and Minimum Potential Niche models assume that the niche ordering (usually thought to represent 
 body size, possibly in combination with other biological features) to some extent constrains species to find prey within closed 
 intervals thereof. A bias could likewise be introduced in the Boltzmann factor of the PPM such that intervals of the sequence of entry 
 were preferred, if this constraint in empirical networks turned out to be more than a spurious effect of trophic coherence.
 
\end{itemize}


\subsubsection{Negative temperatures}

\begin{figure}[ht!]
\renewcommand{\figurename}{Figure S}
\begin{center}
\includegraphics[scale=0.35]{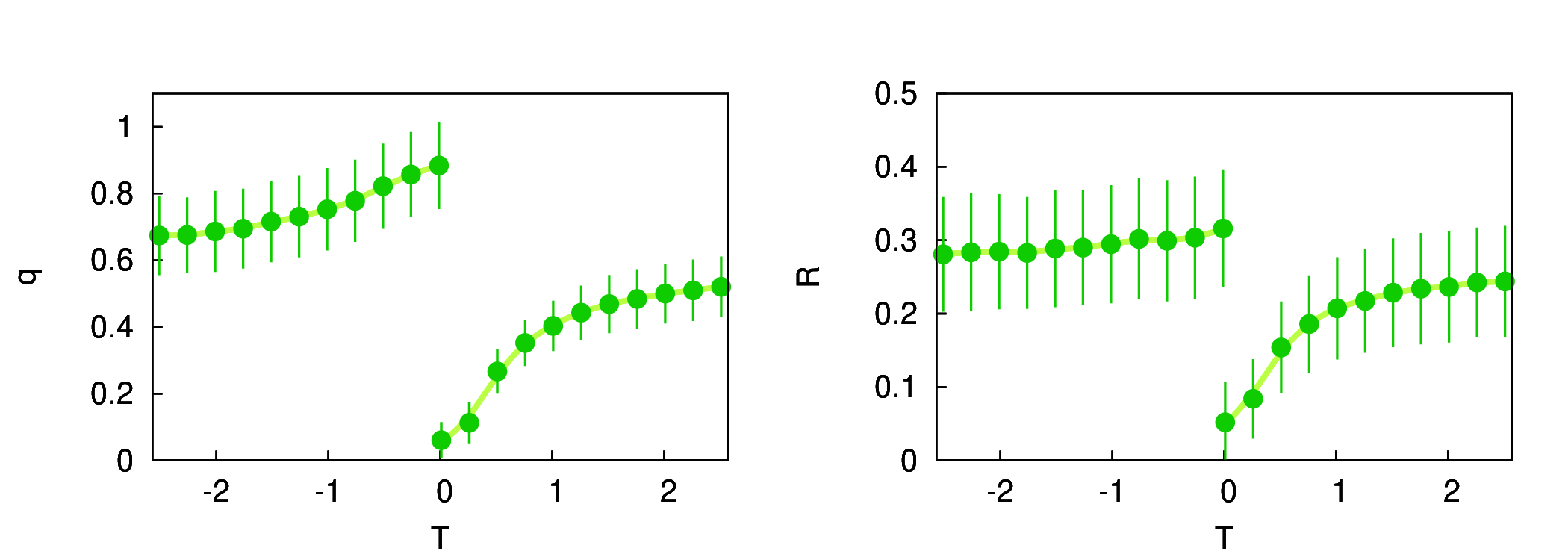}
\end{center}
\caption{
Left: Trophic coherence, as measured by $q$, of networks generated with the PPM with the parameters of 
Chesapeake Bay \cite{chesapeake1,chesapeake2}, 
against $T$, for a range which includes $T<0$. Right: Stability, as measured by $R$, for the 
networks of the panel on the left.
}
\label{fig_Tneg}
\end{figure}

As discussed in the main text, the PPM can generate any level of trophic coherence between that of a maximally coherent structure (with 
$T\rightarrow 0$) and one as incoherent as would be obtained if attachment were random (at $T\rightarrow\infty$). However, as shown 
in Table S\ref{table_foodwebs}, some food webs (five out of the 46 in our dataset) exhibit higher values of $q$ even than this latter 
case. The PPM can also generate greater incoherence than obtained at high positive $T$ with negative values of this parameter, 
as illustrated in Fig. S\ref{fig_Tneg}. The curves of $q$ and $R$ would be continuous if instead of $T$
we used its inverse, $\beta=1/T$. With this parameter, $\beta=0$ corresponds to random attachment, with $q$ falling monotonically from 
maximum incoherence at $\beta\rightarrow -\infty$ to maximum coherence at $\beta\rightarrow +\infty$. A comparison of the two panels in
Fig. S\ref{fig_Tneg} shows that the effect of trophic coherence on stability seems to saturate at about the $q$ obtained with 
random attachment: greater incoherence has little effect on $R$.

\newpage

\section{Food-web data}

\label{Section_Data}

We have compiled a dataset of 46 food webs available in the literature, pertaining to several ecosystem types. The methods used 
by the researchers to establish the links between species vary from gut content analysis to inferences about the behaviour of similar 
creatures. In Table S\ref{table_foodwebs} we list the food webs used along with references to the relevant work. We also list, for each 
case, the number of species $S$, of basal species $B$, the mean degree $K$, the ecosystem type, the trophic coherence $q$, the value of 
the parameter $T$ found to yield (on average) the empirical $q$ with the Preferential Preying Model, and the numerical label 
used to represent the food web in several figures below.

\renewcommand{\tablename}{Table S} 
\setcounter{table}{0}

\begin{center}
 \begin{longtable}{@{}llccccccc}
\hline
  Food web & $S$ & $B$ & $K$ & Type & $q$ & $T$ & Reference & Label\\
\hline
Akatore Stream & 84 & 43 & 2.70 & River & 0.16 & 0.26 & \cite {streams5,streams6,streams7} & 18 \\
Benguela Current & 29 & 2 & 7.00 & Marine & 0.76 & 0.87 & \cite{benguela} & 11\\
Berwick Stream & 77 & 35 & 3.12 & River & 0.18 & 0.25 & \cite {streams5,streams6,streams7} &34\\
Blackrock Stream & 86 & 49 & 4.36 & River & 0.19 & 0.25 & \cite {streams5,streams6,streams7} &27\\
Bridge Brook Lake & 25 & 8 & 4.28 & Lake & 0.59 & 1.15 & \cite{bridge} &14 \\
Broad Stream & 94 & 53 & 6.01 & River & 0.16 & 0.16 & \cite {streams5,streams6,streams7} &35 \\
Canton Creek & 102 & 54 & 6.83 & River & 0.16 & 0.18 & \cite{canton} &2 \\
Caribbean (2005) & 249 & 5 & 13.31 & Marine & 0.75 & 0.70 & \cite{caribbean_2005} &17 \\
Caribbean Reef & 50 & 3 & 11.12 & Marine & 0.99 & -0.24 & \cite{reef} &13 \\
Carpinteria Salt Marsh Reserve & 126 & 50 & 4.29 & Marine & 0.65 & -8.27 & \cite{carpinteria} &33  \\
Catlins Stream & 48 & 14 & 2.29 & River & 0.20 & 0.27 & \cite {streams5,streams6,streams7} &19 \\
Chesapeake Bay & 31 & 5 & 2.19 & Marine & 0.47 & 0.67 & \cite{chesapeake1,chesapeake2} &5 \\
Coachella Valley & 29 & 3 & 9.03 & Terrestrial & 1.34 & -0.02 & \cite{coachella} &12 \\
Crystal Lake (Delta) & 19 & 3 & 1.74 & Lake & 0.28 & 0.33 & \cite{CrystalD} &37 \\
Cypress (Wet Season) & 64 & 12 & 6.86 & Terrestrial & 0.63 & 0.73 & \cite{south_florida98} &42 \\
Dempsters Stream (Autumn) & 83 & 46 & 5.00 & River & 0.23 & 0.30 & \cite {streams5,streams6,streams7} &36 \\
El Verde Rainforest & 155 & 28 & 9.74 & Terrestrial & 1.02 & -0.82 & \cite{el_verde} &15 \\
Everglades Graminoid Marshes & 63 & 5 & 9.79 & Terrestrial & 0.66 & 0.47 & \cite{Everglades} &44 \\
Florida Bay & 121 & 14 & 14.60 & Marine & 0.59 & 0.48 & \cite{south_florida98} &26 \\
German Stream & 84 & 48 & 4.20 & River & 0.21 & 0.29 & \cite {streams5,streams6,streams7} &28 \\
Grassland (U.K) & 61 & 8 & 1.59 & River & 0.40 & 0.72 & \cite{grass} &4 \\
Healy Stream & 96 & 47 & 6.60 & River & 0.22 & 0.24 & \cite {streams5,streams6,streams7} &29 \\
Kyeburn Stream & 98 & 58 & 6.42 & River & 0.18 & 0.18 & \cite {streams5,streams6,streams7} &30  \\
LilKyeburn Stream & 78 & 42 & 4.81 & River & 0.23 & 0.29 & \cite {streams5,streams6,streams7} &31 \\
Little Rock Lake & 92 & 12 & 10.84 & Lake & 0.69 & 0.75 & \cite{little_rock} &8 \\
Lough Hyne & 349 & 49 & 14.66 & Lake & 0.62 & 0.66 & \cite{lough_hyne_1,lough_hyne_2} & 46 \\
Mangrove Estuary (Wet Season) & 90 & 6 & 12.79 & Marine & 0.67 & 0.47 & \cite{south_florida98} &43  \\
Martins Stream & 105 & 48 & 3.27 & River & 0.32 & 0.49 & \cite {streams5,streams6,streams7}&20  \\
Maspalomas pond & 18 & 8 & 1.33 & Lake & 0.48 & -9.22 & \cite{Maspalomas} &39 \\
Michigan Lake & 33 & 5 & 3.91 & Lake & 0.38 & 0.21 & \cite{Michigan} &40 \\
Mondego Estuary & 42 & 12 & 6.64 & Marine & 0.74 & 10.07 & \cite{Mondego} &41 \\
Narragansett Bay & 31 & 5 & 3.65 & Marine & 0.66 & 1.18 & \cite{Narragan} &38 \\
Narrowdale Stream & 71 & 28 & 2.18 & River & 0.25 & 0.38 & \cite {streams5,streams6,streams7} &21  \\
N.E. Shelf & 79 & 2 & 17.76 & Marine & 0.82 & 0.67 & \cite{shelf} &10 \\
North Col Stream & 78 & 25 & 3.09 & River & 0.28 & 0.34 & \cite {streams5,streams6,streams7} &22  \\
Powder Stream & 78 & 32 & 3.44 & River & 0.22 & 0.28 & \cite {streams5,streams6,streams7} &23 \\
Scotch Broom & 85 & 1 & 2.62 & Terrestrial & 0.45 & 0.49 & \cite{broom} &16 \\
Skipwith Pond & 25 & 1 & 7.88 & Lake & 0.68 & 0.23 & \cite{skipwith} &6 \\
St. Marks Estuary & 48 & 6 & 4.60 & Marine & 0.69 & 1.02 & \cite{st_marks} &9 \\
St. Martin Island & 42 & 6 & 4.88 & Terrestrial & 0.59 & 0.60 & \cite{st_martin} &7 \\
Stony Stream & 109 & 61 & 7.61 & River & 0.17 & 0.18 & \cite{stony} &3 \\
Sutton Stream (Autum) & 80 & 49 & 4.19 & River & 0.15 & 0.19 & \cite {streams5,streams6,streams7} &32 \\
Troy Stream & 77 & 40 & 2.35 & River & 0.18 & 0.30 & \cite {streams5,streams6,streams7} &24 \\
Venlaw Stream & 66 & 30 & 2.83 & River & 0.23 & 0.33 & \cite {streams5,streams6,streams7} &25  \\
Weddell Sea & 483 & 61 & 31.81 & Marine & 0.75 & 1.01 & \cite{weddell_sea} & 45 \\
Ythan Estuary & 82 & 5 & 4.82 & Marine & 0.46 & 0.38 & \cite{Ythan96} &1 \\
\caption{
Details of the 46 food webs used throughout the paper. From left to right, the columns are for:
name, number of species $S$, number of basal species $B$, mean degree $K$, ecosystem type, trophic coherence $q$, value of 
the parameter $T$ found to yield (on average) the empirical $q$ with the Preferential Preying Model, references to original work, 
and the numerical label. 
}
\label{table_foodwebs}
\label{table_cascade}
 \end{longtable}
\end{center}

\newpage

\section{Network measures}
\label{Section_Measures}

\subsection{Trophic coherence}
\label{Sec_q}

\begin{figure}[ht!]
\renewcommand{\figurename}{Figure S}
\begin{center}
\includegraphics[scale=0.25]{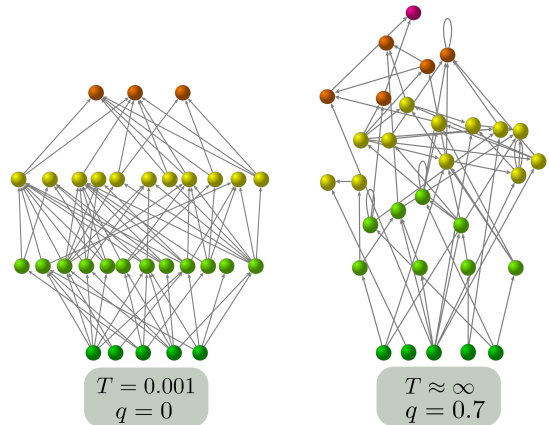}
\end{center}
\caption{
Two example networks generated with the Preferential Preying Model,
illustrating the extremes of trophic coherence: the network on the left was 
generated with $T=0.001$ and has $q=0$ (all links are between species exactly one trophic level apart) while the one on the right
is for $T=10$ (almost random attachment) and has $q=0.7$.
}
\label{fig_visual}
\end{figure}

In the Methods section of the main text we define the network structural property of trophic coherence. Here we simply illustrate 
the difference between a maximally coherent network and a highly incoherent one in Fig. S\ref{fig_visual}.

\begin{figure}[ht!]
\renewcommand{\figurename}{Figure S}
\begin{center}
\includegraphics[scale=0.5, angle=270]{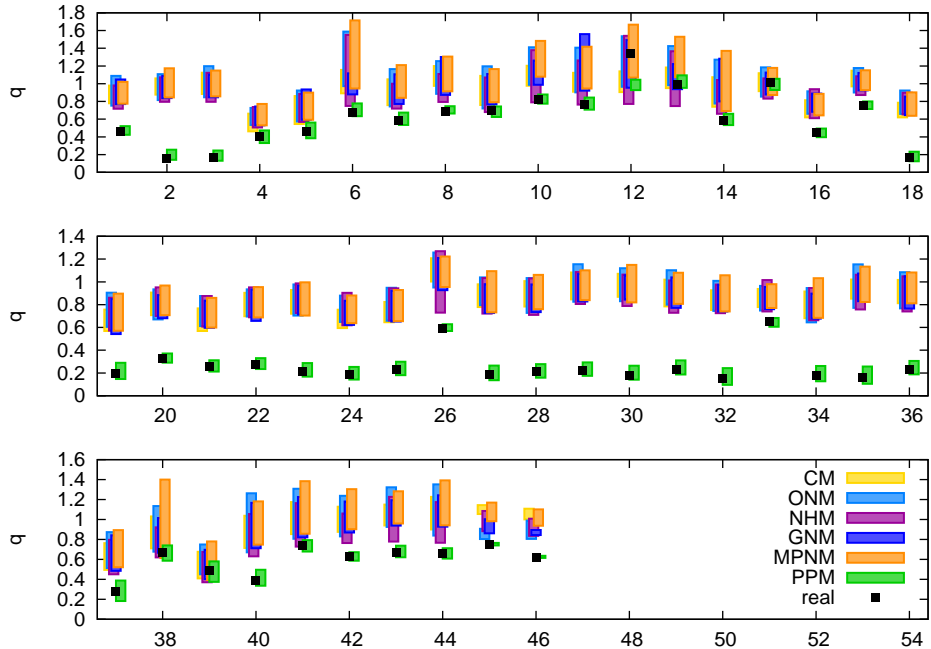}
\end{center}
\caption{
Trophic coherence, as measured by $q$, 
for each of the food webs listed in Table S\ref{table_foodwebs}. The corresponding predictions of each 
food-web model discussed in Section S\ref{Section_Models} -- Cascade, Niche, Nested Hierarchy, Generalized Niche, Minimum Potential Niche
and Preferential Preying -- are displayed with bars representing one standard deviation about the mean. Empirical values are 
black squares. The labelling of the food webs is indicated in the rightmost column of Table S\ref{table_foodwebs}.
}
\label{FigSI_q}
\end{figure}

In Fig. S\ref{FigSI_q} we show the empirical values of $q$ observed in each of the 46 food webs (also displayed in 
Table S\ref{table_foodwebs}) along with the predictions of each of the food-web models discussed above and in the main text.

\subsection{Stability}
\label{sec_stability}

Let us assume that we have a set of ordinary differential equations governing the evolution of the population of each 
species in an ecosystem, as measured, for instance, by its total biomass $x_i$. In vector form, we can write this as
$$
\frac{d}{dt}{\bf x}={\bf f}({\bf x}).
$$
The dynamics will have a fixed point at any configuration ${\bf x^*}$ such that ${\bf f}({\bf x^*})={\bf 0}$. Let us 
suppose that the system is placed at this fixed point but suffers a small perturbation ${\pmb \zeta}(t)$:
$$
{\bf x}(t)={\bf x^*}+{\pmb\zeta}(t).
$$
For small enough $|{\pmb\zeta}(t)|$, its dynamics will be given by the linearised equation:
$$
\frac{d}{dt}{\pmb\zeta}(t)=J({\bf x^*}){\pmb \zeta}(t),
$$
where $J({\pmb x^*})$ is the Jacobian 
matrix $[\partial f_i/\partial x_j]$ evaluated at ${\pmb x^*}$. The fixed point will be 
locally stable if all the eigenvalues of $J({\pmb x^*})$ have negative real part \cite{Stability_scholarpedia}.

Let us consider a fairly general dynamics for ${\pmb x^*}$ given by a consumer-resource model:
\begin{equation}
 \frac{d}{dt}x_i= \eta_{ij}\sum_j a_{ij}F(x_i,x_j) - \sum_j a_{ji}F(x_j,x_i) + G(x_i).
 \label{eq_dynamics}
\end{equation}
The first term on the right accounts for the increment in species $i$'s biomass through consumption of its resources, the second
term is the biomass lost to its consumers, and the function $G$ represents any factors which are not due to interaction with other 
species.
Since we are interested here in effects of interactions between species, we shall simply assume 
$G(x)=\gamma x$ with $\gamma$ a constant.
The function $F$ describes how the interaction between a consumer and a 
resource species depends on their respective biomasses.
The parameter $\eta$ is the efficiency of predation -- the proportion of biomass
lost by a resource which goes on to form part of the consumer.
We shall in general consider this parameter to be constant for all pairs of species ($\eta_{ij}=\eta$, $\forall i,j$), but in 
Sections S\ref{sec_efficiency} and S\ref{sec_weighted} we look into the effects of varying its value.
In the main text, we set this parameter to $\eta=0.2$.

The Jacobian, $J$, will be obtained by taking the partial derivatives of
Eq. (\ref{eq_dynamics}), for each $i$, with respect to each $x_j$.

In the simple case where the interaction between species is given by a sum,
$$
F(x_i,x_j)=x_i+x_j,
$$
we have
\begin{equation}
J_{ij}=(\eta a_{ij} - a_{ji})(1+\delta_{ij})+\gamma\delta_{ij},
\nonumber
\end{equation}
where $\delta_{ij}$ is the Kronecker delta (equal to one when $i=j$, or else zero).
Positive terms added to or subtracted from the main 
diagonal of $J$ simply shift its spectrum of eigenvalues to the right or left, respectively. Therefore, we concentrate on the matrix
\begin{equation}
W=\eta A-A^T,
\label{eq_W}
\end{equation}
where $A^T$ is the transpose of $A$, and consider $\lambda_1$, the eigenvalue of $W$ with the largest 
real part. Then, $R=Re(\lambda_1)$ can be regarded as a measure of the minimum degree of self-regulation at each node which this 
dynamics would require in order for the system to be stable. In other words, the smaller $R$, the more stable we shall say the system 
is.

In this simple case defined by $F(x_i,x_j)=x_i+x_j$ the Jacobian is independent of the point $\pmb x^*$ where it is evaluated.
However, this will not, in general, be the case and for other dynamics we would need to specify this point in order to characterise
the stability of the system. For instance, in a generalised Lotka-Volterra dynamics, the interaction is proportional to the biomass of 
both consumer and resource,
\begin{equation}
F(x_i,x_j)=x_i x_j,
\nonumber
\end{equation}
and the Jacobian becomes
\begin{equation}
J_{ij}=(1+\delta_{ij})w_{ij}x_i+\gamma\delta_{ij},
\label{eq_LV}
\end{equation}
where $w_{ij}$ are the elements of the matrix $W$ as given by Eq. (\ref{eq_W}). Note that this expression depends on the biomass 
of species $i$ (though not on $j$'s) at the point of interest. 

To capture the nonlinearities expected in a prey species' functional response, consumer-resource models often 
describe the interaction as
\begin{equation}
F(x_i,x_j)=x_i H(x_j),
\label{eq_response}
\nonumber
\end{equation}
where $H$ is the Hill equation,
\begin{equation}
 H(x)=\frac{x^h}{x_0^h+x^h},
 \nonumber
\end{equation}
with $x_0$ the 
half-saturation density. The Hill coefficient $h$ determines whether the functional response is of type II ($h=1$) or 
type III ($h=2$) \cite{Real}.
Now we find that the Jacobian is
\begin{equation}
J_{ij}=[\tilde{\eta}(x_i,x_j)a_{ij}-a_{ji}]H(x_i)
\label{eq_Jgen}
\end{equation}
if $i\neq j$, 
where the effective efficiency of predation is
\begin{equation}
\tilde{\eta}(x_i,x_j)=  \frac{x_i}{H(x_i)}\frac{\partial H(x_j)}{\partial x_j}\eta= 
\frac{h x_0^h x_i}{x_j^{h+1}}\frac{H(x_j)^2}{H(x_i)}\eta,
\nonumber
 \end{equation}
and, for the main diagonal elements,
\begin{equation}
J_{ii}= \lbrace h [1-H(x_i)]+1\rbrace H(x_i) w_{ii} +\gamma.
\nonumber
\end{equation}
In each of these kinds of dynamics it is necessary to evaluate the Jacobian at a particular point:
Equations (\ref{eq_LV}) (Lotka-Volterra) and (\ref{eq_Jgen}) (types II and III) are similar in form 
to the matrix $W$ of Eq. (\ref{eq_W}), but their terms are modified by the biomass of the predator, or the biomasses of both prey and 
predator, respectively. One might suggest that we only need identify a fixed point and evaluate the equations there. But, in general, a 
feasible fixed point (in which $x_i>0$ for all $i$) will not exist. 
Feasible fixed points could be defined by 
attributing weights to the elements of the interaction 
matrix $A$, but this would involve decisions on how to do this in a realistic way which might render the results somewhat 
arbitrary. (For a discussion on the feasibility of fixed points, see Section S\ref{sec_feasibility}.)

Throughout most of the paper we focus simply on the matrix $W$ as given by Eq. (\ref{eq_W}), for although the dynamics it describes 
exactly is not very 
realistic (corresponding to the interaction term $F(x_i, x_j)= x_i + x_j$ in Eq. (\ref{eq_dynamics})), it captures the essential 
behaviour of better motivated dynamics without requiring any assumptions about the fixed point.
In fact, if all species had the same biomass at the fixed-point, 
then Eqs. (\ref{eq_LV}) (Lotka-Volterra) and (\ref{eq_Jgen}) (types II and III) would also reduce to the matrix $W$ as given 
by Eq. (\ref{eq_W}), for an appropriate choice of the parameter $\eta$.
However, so as to test the robustness 
of our results to details of the dynamics, in Section S\ref{sec_biomass} we look into the effects of different 
distributions of biomass according together with Lotka-Volterra, type II or type III dynamics.
We find that the relationship between trophic coherence and stability reported in the main text is robust to these considerations, 
although the dependence of biomass on trophic level introduces interesting effects, in particular for the complexity-stability scaling.

\begin{figure}[ht!]
\renewcommand{\figurename}{Figure S}
\begin{center}
\includegraphics[scale=0.5, angle=270]{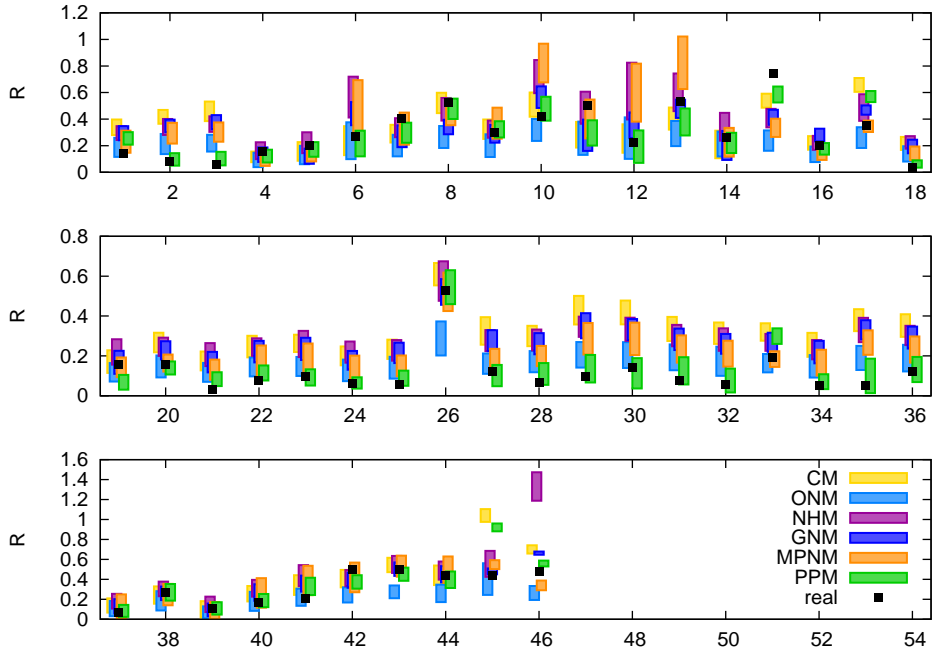}
\end{center}
\caption{
Stability, as measured by $R$, for each of the food webs listed in Table S\ref{table_foodwebs}. The corresponding predictions of each 
food-web model discussed in Section S\ref{Section_Models} -- Cascade, Niche,  Nested Hierarchy, Generalized Niche, Minimum Potential Niche
and Preferential Preying -- are displayed with bars representing one standard deviation about the mean. Empirical values are 
black squares. The labelling of the food webs is indicated in the rightmost column of Table S\ref{table_foodwebs}.
}
\label{FigSI_R}
\end{figure}

\begin{figure}[ht!]
\renewcommand{\figurename}{Figure S}
\begin{center}
\includegraphics[scale=0.5, angle=270]{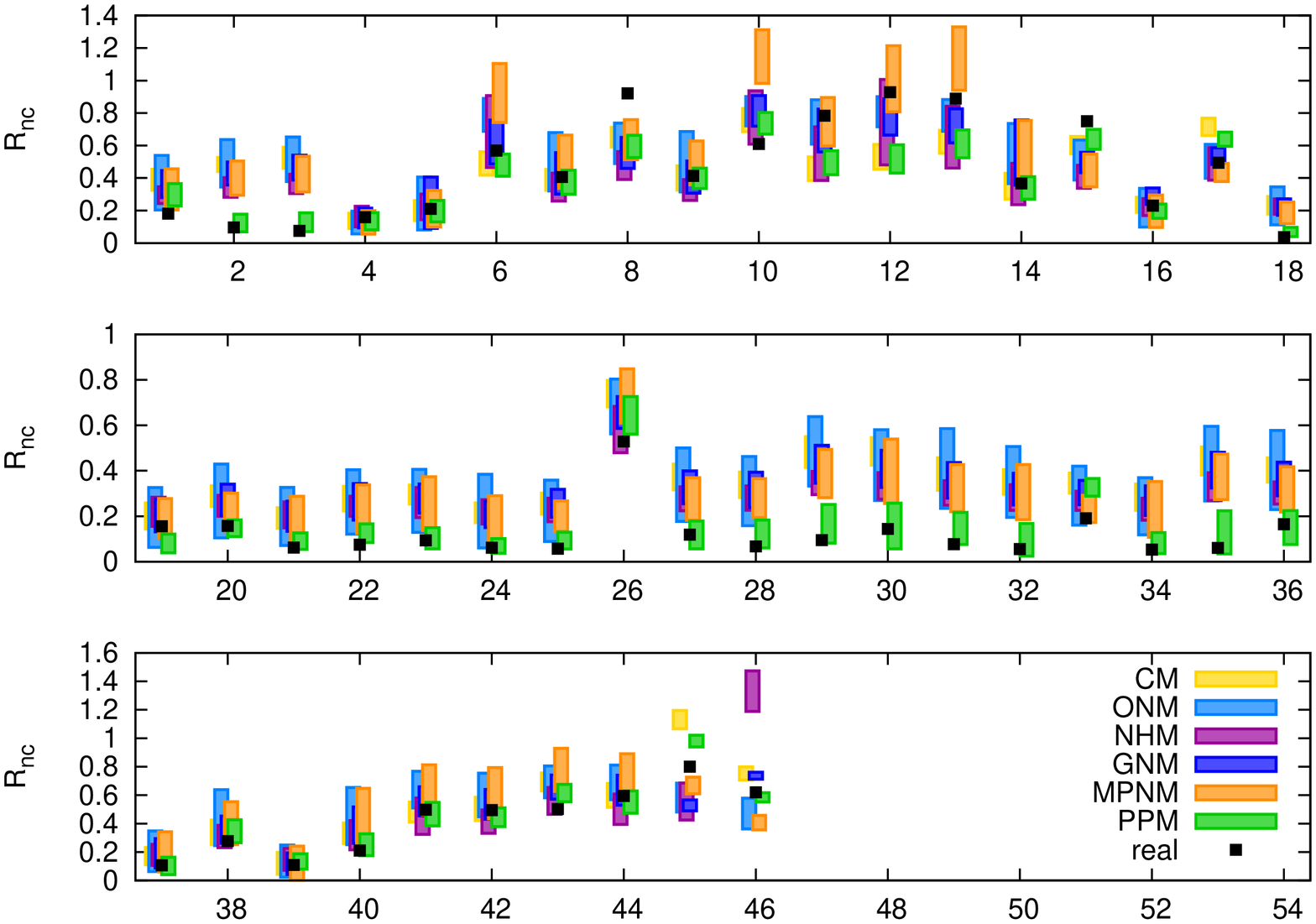}
\end{center}
\caption{
Stability after removal of all self-links, $R_{nc}$,
as measured by $R$, for each of the food webs listed in Table S\ref{table_foodwebs}. The corresponding predictions of each 
food-web model discussed in Section S\ref{Section_Models} -- Cascade, Niche, Nested Hierarchy, Generalized Niche, Minimum Potential Niche
and Preferential Preying -- are displayed with bars representing one standard deviation about the mean. Empirical values are 
black squares. The labelling of the food webs is indicated in the rightmost column of Table S\ref{table_foodwebs}.
}
\label{FigSI_Rnc}
\end{figure}

In the main text we describe how stability in directed networks (and food webs in particular) is determined to a large extent by their
trophic coherence. In Fig. S\ref{FigSI_R} we compare the predictions of each of the food-web models described in 
Section \ref{Section_Models} for each of the food webs listed in Table \ref{table_foodwebs}.
Another network feature which influences stability, as mentioned above, is the existence of self-links 
(representing cannibalism, in the case of food webs),
since this is a form of self-regulation. We disentangle this effect from that of trophic coherence, we remove all self-links from the 
food webs and again measure the real part of the leading eigenvalue, $R_{nc}$. The predictions of each model are shown in 
Fig. S\ref{FigSI_Rnc}.

In Section \ref{sec_luca} we give a proof that a maximally coherent network ($q=0$) with constant interaction strengths
can always be stabilised with an infinitesimal degree of self-regulation.

\subsubsection{Biomass distribution}
\label{sec_biomass}

As discussed in Section S\ref{sec_stability}, the Jacobian corresponding to most kinds of biologically plausible dynamics will depend
on details of the fixed point. In other words, we need to know the biomass of each species in order to evaluate the Jacobian.
Since only a fraction of the energy produced by a species can be used by its consumers, ecosystems can often be regarded as pyramids in 
which biomass is a decreasing function of trophic level \cite{Agrawal}. More specifically, if we assume that the biomass of a species 
is a constant fraction of the combined biomass of its resources, biomass will be exponentially related to trophic level.
We can thus write
\begin{equation}
x_i=e^{a(s_i-1)},
\label{eq_xi_nonoise}
\end{equation}
with $a$ a parameter determining the difference in biomass between predator and prey species (for $a=0$ there is no dependence
of biomass on trophic level), and set the basal species to unity biomass.
A negative value of $a$ then corresponds to a pyramid in which biomass decreases with trophic level (note that a graphical representation 
of this situation will look like a pyramid if the size of each echelon corresponds to the logarithm of its biomass).
Although terrestrial food webs have this distribution, in certain aquatic environments inverted pyramids can arise, corresponding to a 
positive $a$. This is due to the effect of increasing longevity with trophic level, which can compensate to some extent
for the inefficiency of predation \cite{Agrawal}.

In order to examine the robustness of results to fluctuations in this exponential law, we can consider instead a biomass given by
\begin{equation}
x_i=(1+\xi_i)e^{a(s_i-1)},
\label{eq_xi}
\end{equation}
where the variables $\xi_i$ are randomly drawn from a normal distribution with mean zero and standard deviation $\sigma_x$.
We can then use these values of ${\pmb x}$ to evaluate the Jacobian for each kind of dynamics and study the behaviour of its 
leading eigenvalue, $R$.

\begin{center}
 \begin{longtable}{@{}c c|c|c|l}
Jacobian & $\sqrt{S} $ & $\sqrt{K}$ & $q$ & $q$ (no self-links) \\
\hline
\hline
$W$ & $0.064$ & $0.461$  & $0.596$ & $0.804$ \\
\hline
$W_I$& $0.045$ & $0.219$ & $0.431$ & $0.730$ \\
\hline
$W_{II}$ & $0.088$ & $0.359$ & $0.456$ & $0.658$ \\
\hline
$W_{III}$ & $0.107$ & $0.426$ & $0.608$ & $0.582$ \\
\hline
\caption{
First column: Jacobian used to compute stability of the empirical food webs of Table S\ref{table_foodwebs}. 
$W$ is simply the interaction matrix, as used throughout the main text; $W_I$, $W_{II}$ and $W_{III}$ correspond 
to types I, II and III, respectively (where Lotka-Volterra is type I). For these cases, we assume an uncorrupted biomass pyramid,
as given by $a=-0.2$ in Eq. (\ref{eq_xi_nonoise}).
Second, third and fourth column, respectively: value of the correlation coefficient $r^{2}$
obtained for $R$ (stability) against $\sqrt{S}$ (where $S$ is the number of species), $\sqrt{K}$ (where $K$ is the mean degree),
and $q$ (incoherence parameter). Fourth column: as the third column, after removing all self-links.
Compare with Fig. 1 of the main text.}
\label{table_TYPE}
\end{longtable}
\end{center}

Table S\ref{table_TYPE} shows the correlations between stability and the various network measures shown in Fig. 1 of the main 
text over the 46 food webs in the dataset. The first row displays the values for the simple case where the Jacobian is 
considered equal to the interaction matrix $W$. The second, third and fourth rows are for the cases of Lotka-Volterra, type II and 
type III dynamics, with biomass distributed according to Eq. (\ref{eq_xi_nonoise}) and $a=-0.2$. The general pattern shown in
Fig. 1 of the main text is conserved for these more realistic dynamics.

\begin{figure}[ht!]
\renewcommand{\figurename}{Figure S}
\begin{center}
\includegraphics[scale=0.35]{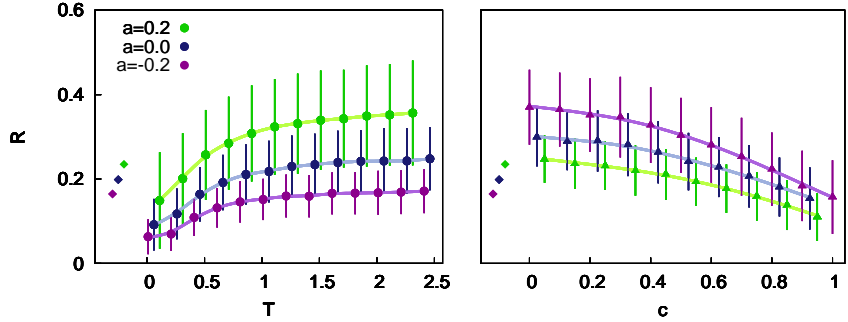}
\end{center}
\caption{
Value of $R$ obtained for the Lotka-Volterra Jacobian given by Eq. (\ref{eq_LV}), with biomass distributed according to 
Eq. (\ref{eq_xi_nonoise}) for $a=-0.2$ (pyramid), $0$ (flat), and $0.2$ (inverted pyramid).
In each panel, the diamonds represent the 
values for the empirical food web of Chesapeake Bay \cite{chesapeake1,chesapeake2}.
Circles in the panel on the left show the corresponding results for PPM networks against $T$
using the same parameters; triangles in the panel on the right are for networks generated with the Generalized Niche Model 
against contiguity, $c$.
}
\label{fig_pyramid}
\end{figure}

In Fig. \ref{fig_pyramid} we show the values of $R$ obtained from the Lotka-Volterra Jacobian given by Eq. (\ref{eq_xi_nonoise})
with different values of $a$, corresponding to pyramid, flat and inverted pyramid distributions of biomass.
The empirical values 
found for the Chesapeake Bay food web \cite{chesapeake1,chesapeake2} with each distribution are compared to the predictions of 
the Preferential Preying Model against $T$ (left panel), and the Generalized Niche Model against contiguity $c$ (right panel).
The effect of the parameter $T$ on stability in the PPM networks remains qualitatively the same as the results reported in the 
main text for the matrix $W$ given by Eq. (\ref{eq_W}). The more squat the biomass pyramid (the more negative the parameter $a$),
the more stable are both the empirical and PPM networks. This is in keeping with observations of ecosystems \cite{Agrawal}.
In the Generalized Niche Model networks, however, the effect is opposite: it is the inverted pyramid (positive $a$),
which is most stable. We do not have an explanation for such an effect, but note that it marks a qualitative 
difference between the networks generated with this model and real food webs.

\begin{figure}[ht!]
\renewcommand{\figurename}{Figure S}
\begin{center}
\includegraphics[scale=0.63]{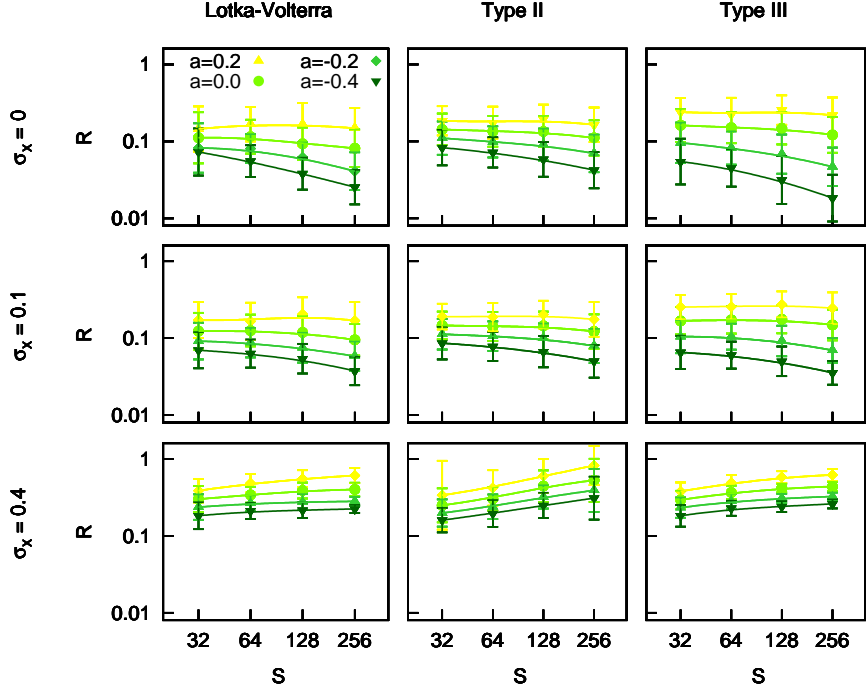}
\end{center}
\caption{
Scaling of $R$ with $S$ in PPM networks generated with $T=0.01$, $K=S^{0.4}$, and $B=0.25 S$.
In each panel, from top to bottom, lines are for $a=0.2$, $0$, $-0.2$ and $-0.4$.
From left to right, columns of panels are 
for Lotka-Volterra, type II and type III dynamics, as given by Eqs. (\ref{eq_LV}) and (\ref{eq_Jgen}).
From top to bottom, rows of panels are 
for levels of biomass noise $\sigma_x=0$, $0.1$ and $0.4$ in Eq. (\ref{eq_xi}).
In types II and III,
the half-saturation is set at $x_0=1/2$.
}
\label{fig_multi}
\end{figure}

In Fig. \ref{fig_multi} we look into how the biomass distribution affects the diversity-stability relationship.
All networks are generated with the Preferential Preying Model and
$T=0.01$.
The first row of panels is for the case where biomass decays with trophic level as an uncorrupted exponential ($\sigma_x=0$),
for Lotka-Volterra, type II and type III dynamics (top panels from left to right). As compared with the constant biomass case 
($a=0$), a decaying distribution is seen to increase the slope whereby $R$ falls with $S$. In other words, placing more 
biomass at the bottom of the food web than at the top not only increases stability, 
but also strengthens the 
positive diversity-stability relationship exhibited by trophically coherent networks.
This occurs for all three kinds of dynamics, although the effect is strongest for type III and weakest for type II.
For an inverted pyramid (positive $a$), $R$ is approximately constant with $S$.

We go on to analyse the effect of corrupting the exponential distribution of biomass with a noise of standard deviation $\sigma_x$.
The second row of panels is for $\sigma_x=0.1$.
Although the slope is now less pronounced in all cases, this degree of noise does not undermine the positive 
diversity-stability relationship for any of the dynamics considered.
Finally, in the bottom row we apply a higher noise, $\sigma_x=0.4$. Now the relationship is 
inverted and diversity decreases stability. It is not, perhaps, surprising that noise in the distribution of biomass (large $\sigma_x$) 
should have a similar effect on scaling as incoherence in the trophic structure (large $T$). However, it is interesting that the noise 
level at which the transition from a positive to a negative diversity-stability relationship occurs does not seem to depend on 
$a$ or on the kind of dynamics.

\begin{figure}[ht!]
\renewcommand{\figurename}{Figure S}
\begin{center}
\includegraphics[scale=0.6]{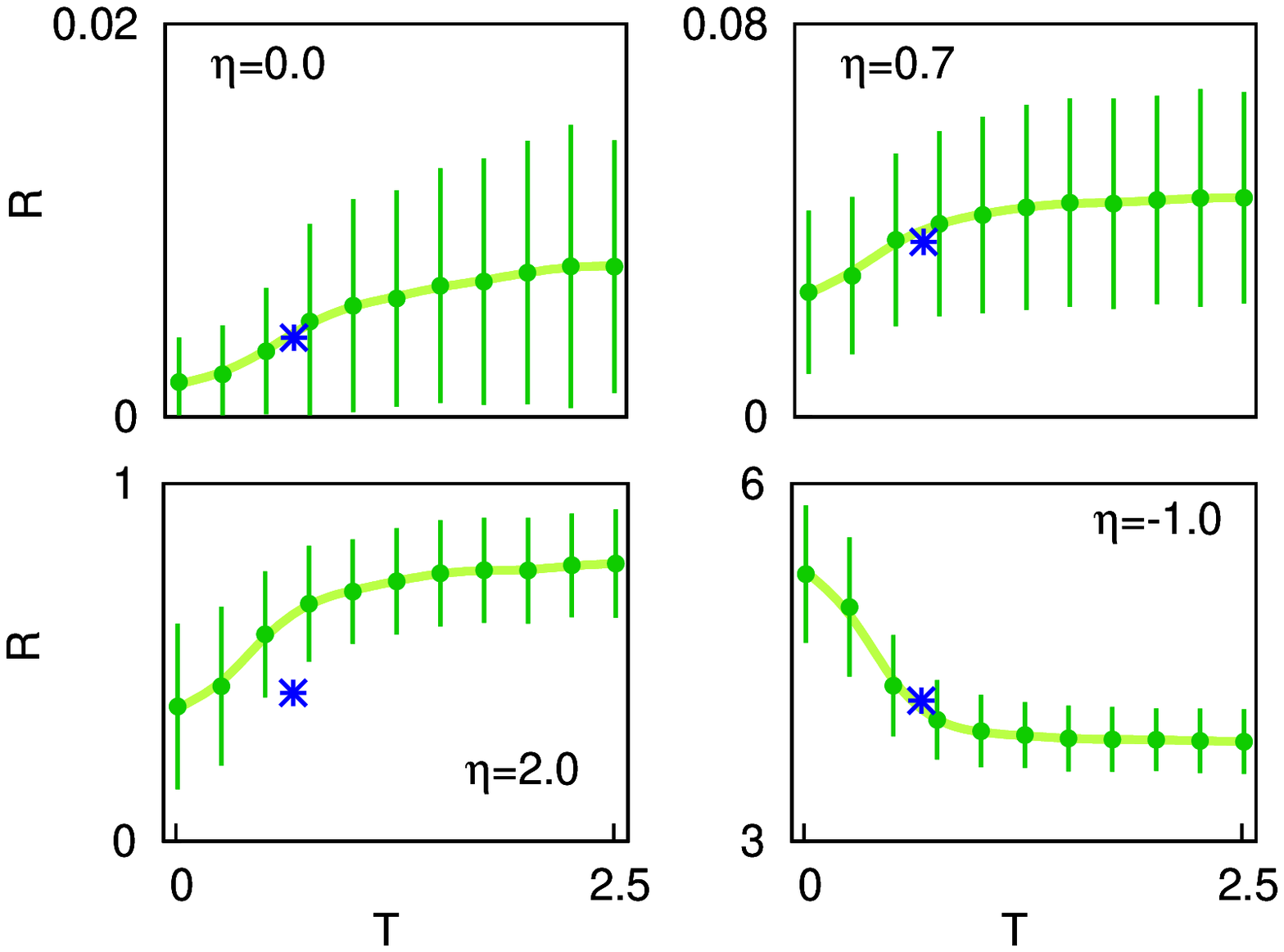}
\end{center}
\caption{
Real part of the leading eigenvalue, $R$, of the interaction matrix $W=\eta A-A^T$ against the parameter $T$, from averages over networks 
generated with the PPM for the parameters of the Chesapeake Bay food web \cite{chesapeake1,chesapeake2}. In each panel a different value of 
the parameter $\eta$ is used, and the corresponding empirical value of $R$ is represented with a blue asterisk at the value $T=0.67$, 
found to predict the empirical trophic coherence $q=0.47$ (as shown in Table S\ref{table_foodwebs}). Top left: $\eta=0$; 
top right: $\eta=0.7$; bottom left $\eta=2$; bottom right $\eta=-1$.
}
\label{FigSI_eta}
\end{figure}

\subsubsection{Efficiency}
\label{sec_efficiency}
According to the definition of $R$ above, we must give a value to the parameter $\eta$ in order to measure 
stability. The value of 
this parameter affects the kind of interaction we intend to model with the interaction matrix, $W=\eta A-A^T$, 
and has a strong bearing on the values of
$R$ measured. The definition of $W$ captures the fact that the effect of a prey species on one of 
its predators is a proportion $\eta$ of the effect of the predator on the prey. If we are considering the flow of biomass from prey to 
predator, this should be a relatively small fraction -- for instance, the ``ten percent law'' is often used as a rough 
estimate of the efficiency of predation \cite{efficiency}. On the other hand, our definition of stability is only strictly independent of 
the fixed point for a dynamics such as the one described above. For a more realistic dynamics, we might expect a multiplicative factor 
to appear relating the fixed-point biomass of a prey species to that of one of its predators. The parameter $\eta$ might therefore 
be increased (or decreased) by this effect.

As mentioned above, throughout the paper we use the value $\eta=0.2$.
However, simulations of the PPM show that using the value of the parameter $T$ which best approximates the empirical 
degree of trophic coherence is enough to predict the empirical $R$ for a wide range of $\eta$. In Fig. S\ref{FigSI_eta} we show $R$
against $T$ for PPM networks constructed with the parameters of the Chesapeake Bay food web \cite{chesapeake1,chesapeake2} for four cases.
We also plot, with an asterisk, the empirical value of $R$ observed in each case, always at the value $T=0.67$ found to adjust the 
empirical trophic coherence, $q=0.47$ (see Table S\ref{table_foodwebs}). The top left panel is for the case of $\eta=0$, which represents 
a situation in which the biomass of prey species is completely unaffected by the biomass of their predators. We show in the proof we 
include in Methods that a perfectly coherent network with $\eta=0$ would have only zero eigenvalues. As incoherence increases, $R$ grows 
somewhat, though it remains small compared to most cases in which the parameter $\eta$ simulates a measure of feedback from predators to 
prey. The top right panel is for $\eta=0.7$, implying a relatively high efficiency and a strong negative feedback acting on 
prey species. At $\eta=1$, all the eigenvalues of $W$ would have zero real part because it would be an antisymmetric matrix (intuitively, 
any increase in one node's biomass will be compensated by a decrease in another, so perturbations will be maintained and neither dampened 
nor amplified). At $\eta>1$ we simulate a situation such that a predator extracts more biomass form its prey than the latter loses. As 
we would expect intuitively, this scenario of runaway growth is significantly more unstable than the ones described above.
However, the behaviour of $R$ with $T$ is qualitatively similar to that observed for $0<\eta<1$. 
Finally, the bottom right panel corresponds to the case $\eta=-1$, implying that predation reduces the biomass of a predator as well
as that of its prey. We know from the proof described in Methods that at $q=0$ all the eigenvalues of $W$ are purely real for 
any $\eta<0$. Similarly, the behaviour of $R$ with $T$ is now inverted: the most coherent networks are now the most unstable.

In the panels corresponding to $\eta=0$, $0.7$ and $-1$, the value of $T$ which adjusts the empirical trophic coherence also predicts 
the empirical $R$ very accurately (as we have found for all the food webs in our dataset when using $\eta=0.2$; see main text). The case 
of $\eta=2$ is slightly out: the PPM predicts a slightly higher value of $R$ at $T=0.67$, although it is not out by much more than a 
standard deviation. This case of $\eta>1$ is unlikely to be relevant for ecology; but the small discrepancy serves to remind us that 
the PPM does not capture all the structural features of real food webs.

\subsubsection{Herbivory}
\label{sec_herbivory}

Links from basal species (producers) to species which only consume basal species (herbivores) will necessarily have 
a trophic distance equal to one (see Methods in the main text). Since the proportion of basal species, $B/S$, varies considerably 
among food webs, we can expect this measure to have a strong bearing on trophic coherence. On the other hand, a large number of 
basal species may provide a more stable configuration than a network in which many species depend on just a few producers. Might 
this be the underlying reason for the relation between trophic coherence and stability?

\begin{figure}[ht!]
\renewcommand{\figurename}{Figure S}
\begin{center}
\includegraphics[scale=0.42]{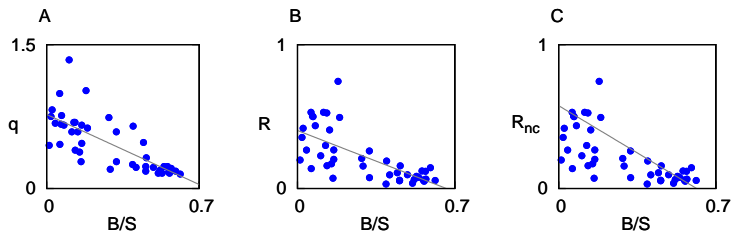}
\end{center}
\caption{
Scatter plots, for the food webs listed in Table S\ref{table_foodwebs}, of three network measures against the proportion of basal 
species, $B/S$, with Pearson's correlation coefficient in brackets.
{\bf A}: Trophic coherence, $q$, against $B/S$ ($r^2=0.559$).
{\bf B}: $R$ (real part of the leading eigenvalue of $W$) against $B/S$ ($r^2=0.475$).
{\bf C}: $R_{nc}$ (real part of the leading eigenvalue of $W$ after self-links have been removed) against $B/S$ ($r^2=0.505$).
}
\label{FigSI_B}
\end{figure}

Figure S\ref{FigSI_B}A is a scatter plot of $q$ against $B/S$ for the food webs listed in Table S\ref{table_foodwebs}. There is indeed a 
significant negative correlation ($r^2=0.559$). Figures S\ref{FigSI_B}B and S\ref{FigSI_B}C show how stability, as 
measured both before and after removing self-links, varies with the proportion of basal species in the same dataset. The correlations 
are also significant ($r^2=0.475$ for $R$ and $r^2=0.505$ for $R_{nc}$), but slightly lower than we observe in Fig. S\ref{FigSI_B}A.
In any case, they are much weaker than the correlations shown in Fig. 1 of the main text between trophic coherence and stability. We can 
therefore conclude that trophic coherence is the most powerful explanatory variable of stability, while the effect of the proportion of 
basal species is either less important, or simply an artefact of its correlation with trophic coherence.

\subsubsection{Weighted networks}
\label{sec_weighted}
\begin{figure}[ht!]
\renewcommand{\figurename}{Figure S}
\begin{center}
\includegraphics[scale=0.35]{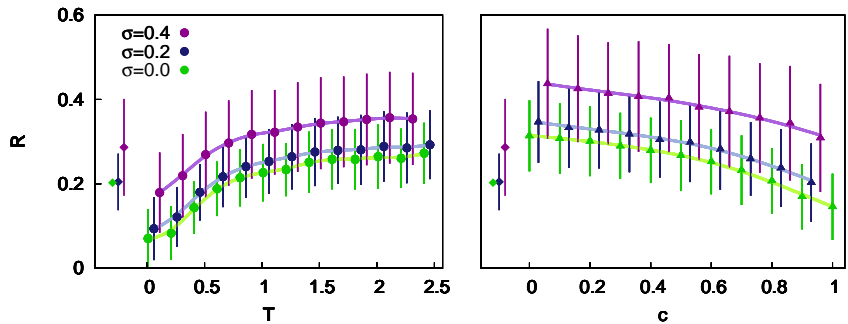}
\end{center}
\caption{
Value of $R$ obtained after defining the modified interaction matrix $\tilde{W}_{ij}= (1+\xi_{ij})W_{ij}$, 
where $\xi_{ij}$ is drawn from a Gaussian distribution of mean zero and standard deviation $\sigma$,
and averaging over realizations of the noise $\lbrace \xi \rbrace$. In each panel, the diamonds represent the 
average values for the empirical food web of Chesapeake Bay \cite{chesapeake1,chesapeake2}, with standard deviations as error bars, 
for noise levels $\sigma=0$, $0.2$ and $0.4$. The panel on the left shows the corresponding results for PPM networks against $T$
using the same parameters, while the panel on the right is for those generated by the Generalized Niche Model against contiguity, $c$.
}
\label{FigSI_noise}
\end{figure}

Although we have been considering the food webs as unweighted networks (the elements in $A$ are either zero or one), in reality 
certain interactions will be more important than others,
and the efficiency $\eta$ need not be the same for all links.
A simple way 
to look into how these considerations might affect our results is as follows. We make the change 
$W_{ij}\rightarrow (1+\xi_{ij})W_{ij}$, with $\xi_{ij}$ drawn from a Gaussian distribution of mean zero, standard 
deviation $\sigma$ and no correlation between $\xi_{ij}$ and $\xi_{ji}$. For a given network we then obtain the value of $R$ for 
many different realizations of the noise $\lbrace \xi \rbrace$.
In the left panel of Fig. S\ref{FigSI_noise}
we show the average and standard deviations of $R$ thus defined for three different levels of noise -- $\sigma=0.0$, $0.2$ and $0.4$ --
for PPM networks with the parameters of the Chesapeake Bay food web \cite{chesapeake1,chesapeake2}. We also show (with diamonds) the 
corresponding averages and standard deviations obtained by performing the same test on the empirical food web. As is to be expected, 
increased noise leads to a higher average $R$ (lower stability) and a wider standard deviation. However, the behaviour of the 
average $R$ against the parameter $T$ remains similar with increasing noise, and the value $T=0.67$ which best 
adjusts the empirical trophic coherence (as given by Table S\ref{table_foodwebs}) continues to predict the empirical average $R$ at each 
$\sigma$. This is not, however, the case for the Generalized Niche Model. We show the mean and standard deviation of $R$ generated with 
this model against its contiguity parameter $c$ for the same food web. Whereas the empirical and simulated average values of $R$
correspond at $c\lesssim1$ when there is little noise, as $\sigma$ increases the model average $R$ grows faster than the 
empirical value. This suggests that trophically coherent networks, such as the Chesapeake Bay food web or those generated by the PPM, 
are more robust to fluctuations in interaction strengths than those generated with niche-based models.

The allometric relationship according to which metabolic rates decline with increasing body size has 
been shown to reduce predation strength per unit biomass, thereby contributing to stability \cite{neodinamic}.
Since body size tends to augment (exponentially) with trophic level, this would mean that a more coherent structure 
would also involve a more homogeneous distribution of link strengths (for a given predator). Therefore, in a more realistic 
setting in which body sizes and link strengths are considered, we expect the stabilising effect of trophic coherence to 
be greater than we have shown here for binary networks.

\subsubsection{Feasibility}
\label{sec_feasibility}

We have been discussing the potential stability of fixed points of ecosystem dynamics, but for this to 
be relevant such a fixed point has to be {\it feasible}. That is, there must exist a fixed point such that every species has a
positive biomass. To determine a potential fixed point one must, in general, know the details of the dynamics (as mentioned above).
However, even with these specifications, given an unweighted network is is highly unlikely that the fixed point will involve only positive
biomasses. However, nature does not have this problem, among other reasons because species' biomasses co-evolve with the interaction 
weights. If we are granted a certain freedom to set these weights, even if other details of dynamics are set, the problem of finding 
a fixed point becomes under-specified, and configurations allowing for feasible fixed points might be located. 
We saw above that the stability of real food webs and 
those generated by the PPM seem to be more robust to random changes in interaction strengths than their niche-based model counterparts.
This suggests that, given a prescription to modify interaction weights, trophic coherence might enhance the feasibility of fixed points 
as well as their stability. Such an exercise lies beyond the scope of this paper, but we believe it is a promising avenue of research 
to be undertaken in the future.

\subsubsection{Stability criteria}

\begin{figure}[ht!]
\renewcommand{\figurename}{Figure S}
\begin{center}
\includegraphics[scale=0.42]{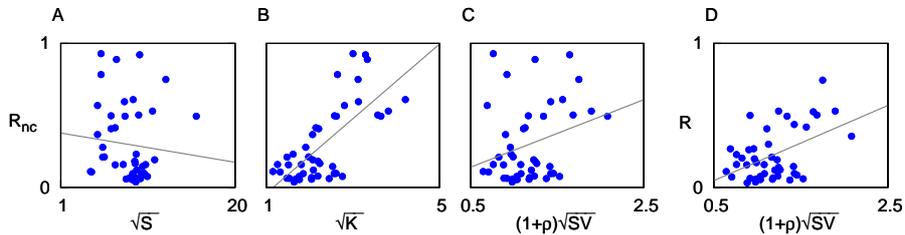}
\end{center}
\caption{
Scatter plots, for the food webs listed in Table S\ref{table_foodwebs}, of stability measures against various network values, 
with Pearson's correlation coefficient in brackets.
{\bf A}: $R_{nc}$ (real part of the leading eigenvalue after self-links have been removed) against $\sqrt{S}$ ($r^2=0.008$).
{\bf B}: $R_{nc}$ against $\sqrt{K}$ ($r^2=0.508$).
{\bf C}: $R_{nc}$ against Allesina and Tang's prediction, given by Eq. (\ref{eq_alle}) ($r^2=0.083$).
{\bf D}: $R$ (real part of the leading eigenvalue without removing self-links) against Allesina and Tang's prediction ($r^2=0.230$).
}
\label{FigSI_alle}
\end{figure}

In the main text we discuss May's result for random networks, according to which the real part of 
the leading eigenvalue should scale as $R\sim \sqrt{SC}=\sqrt{K}$. We also show that $R$ does not exhibit a significant correlation 
with $\sqrt{S}$, although we do observe a modest positive correlation ($r^=0.480$) with $\sqrt{K}$. In Figs. S\ref{FigSI_alle}A and 
S\ref{FigSI_alle}B we show scatter plots, for the food webs listed in Table S\ref{table_foodwebs}, of the leading eigenvalue after 
self-links have been removed, $R_{nc}$, against 
$\sqrt{S}$ and $\sqrt{K}$.
In the former case the correlation is now negative but still insignificant, while in the latter the 
correlation increases slightly to $r^2=0.508$. However, food webs are network in which all the links stand for predation (as opposed 
to other ecological relationships, such as competition or mutualism). Allesina and Tang have recently derived stability criteria 
for specific kinds of interactions \cite{Allesina_Tang}. In particular, when the links stand for predation but are randomly placed among the species, 
they find that the real part of the leading eigenvalue should scale as
\begin{equation}
 R\sim (1+\rho)\sqrt{SV},
\label{eq_alle}
\end{equation}
where $V$ is the variance of the off-diagonal elements of the interaction matrix $W$, and $\rho$ is Pearson's correlation coefficient 
between the elements $W_{ij}$ and $W_{ji}$. Figure \ref{FigSI_alle}C is a scatter plot of $R_{nc}$ against the prediction of 
Eq. (\ref{eq_alle}). Somewhat surprisingly, the correlation is very weak ($r^2=0.083$). In Fig. \ref{FigSI_alle}D we swap $R_{nc}$ for
$R$ (the leading eigenvalue when cannibalism is included) and now the correlation becomes significant ($r^2=0.230$), although 
still relatively low. These results provide further evidence that the structure of food web is non-random in a way which is 
particularly relevant for their stability.

\subsubsection{Missing links and trophic species}
\label{sec_missing}

Despite important recent developments in food-web inference techniques, it is often hard to ascertain from observation whether 
a given species consumes another (and even more difficult to quantify the extent of predation). Furthermore, the food webs 
we have used here for our analysis (described in Section \ref{Section_Data}) were obtained with a variety of different techniques. 
To assess whether the patterns we have 
observed in this dataset, shown in Fig. 1 of the main text, are robust to possible experimental errors, we remove from each food web 
a percentage of links, chosen randomly, and recompute each of the magnitudes of interest. After averaging over 100 such tests for each 
food web, we then recalculate each of the correlation coefficients shown in Fig. 1. These are shown in Table S\ref{table_deletion} 
for 
different percentages of links removed.
As we can see, the dependency of stability on the other magnitudes is barely affected by the random deletion of links:
the correlation of $R$ with size is never significant, while the correlation with both complexity and coherence actually increases 
slightly with the percentage of deleted links.

\begin{center}
 \begin{longtable}{@{}c c|c|c|l}
Missing links & $\sqrt{S} $ & $\sqrt{K}$ & $q$ & $q$ (no self-links) \\
\hline
\hline
$0\%$ & $0.064$ & $0.461$  & $0.596$ & $0.804$ \\
\hline
$1\%$ & $0.061$ & $0.484$ & $0.598$ & $0.814$ \\
\hline
$5\%$ & $0.064$ & $0.497$ & $0.635$ & $0.831$ \\
\hline
$10\%$ & $0.014$ & $0.545$ & $0.752$ & $0.857$ \\
\hline
$20\%$ & $0.002$ & $0.582$ & $0.783$ & $0.845$ \\
\hline
\caption{
First column: percentage of links randomly deleted from the empirical food webs of Table S\ref{table_foodwebs}.
Second, third and fourth column, respectively: value of the correlation coefficient $r^{2}$
obtained for $R$ (stability) against $\sqrt{S}$ (where $S$ is the number of species), $\sqrt{K}$ (where $K$ is the mean degree),
and $q$ (incoherence parameter). Fourth column: as the third column, after removing all self-links.
Compare with Fig. 1 of the main text.}
\label{table_deletion}
\end{longtable}
\end{center}

The nodes in the food webs found in the literature often represent ``trophic species''. 
This means that if two or more species in the community share their 
full sets of prey and predators, they are coalesced into a single node, even if they are in fact taxonomically distinct.
However, with recent advances in empirical techniques of food-web inference, larger networks are now being obtained in 
which nodes represent taxonomic, rather than trophic, species.
To find out whether our empirical findings are affected by the degree of taxonomic resolution, we perform a similar test to that of 
link deletion: for each food web, we randomly choose a percentage of species to be duplicated -- that is, we introduce a new species 
with the same sets of predators and prey. As before, we average over 100 such tests and recalculate the correlation coefficient for 
each pair of magnitudes of interest. In Table S\ref{table_duplicity} we show these results for various percentages of duplicated species.
As with the deleted links, we find that the correlations are fairly robust to these modifications, implying that they are not severely 
affected by the taxonomic resolution of the food webs.

\begin{center}
 \begin{longtable}{@{}c c|c|c|l}
Species duplicated & $\sqrt{S} $ & $\sqrt{K}$ & $q$ & $q$ (no self-links) \\
\hline
\hline
$0\%$ & $0.064$ & $0.461$  & $0.596$ & $0.804$ \\
\hline
$20\%$ & $0.002$ & $0.582$ & $0.783$ & $0.845$ \\
\hline
$50\%$ & $0.122$ & $0.406$ & $0.713$ & $0.797$ \\
\hline
\caption{First column: percentage of species duplicated (as described in Section \ref{sec_missing}) 
in the empirical food webs of Table S\ref{table_foodwebs}.
Second, third and fourth column, respectively: value of the correlation coefficient $r^{2}$
obtained for $R$ (stability) against $\sqrt{S}$ (where $S$ is the number of species), $\sqrt{K}$ (where $K$ is the mean degree),
and $q$ (incoherence parameter). Fourth column: as the third column, after removing all self-links.
Compare with Fig. 1 of the main text.
}
\label{table_duplicity}
\end{longtable}
\end{center}


\subsection{Mean chain length}

A food chain is a directed path
beginning at at basal species (one with no in-coming links) and ending at 
an apex predator (one with no out-going links) \cite{Elton}. In other words, it is any one of the possible paths that 
biomass entering the system through a basal species can follow until it is entirely dissipated. A food web generally has 
a very large number of such chains; but a low mean chain length (MCL) -- an average over all of them, the length of a chain being 
the number of links it comprises -- has been 
associated with a high stability \cite{Pimm}.


All food webs representing a more or less autonomous 
ecosystem necessarily have at least one basal species; however, it can occur that there are in fact no apex predators. 
This is because the top predators can eat each other. To get round this we define an {\it apex set} as a group of 
predators such that no directed paths leave the group, while they would if 
any member of the set were removed. For instance, say predator A and predator B would both be apex in the usual sense 
if it weren't because they ate each other. With this definition they form, together, an apex set.
Thus, we define a food chain as a directed path beginning at a basal species 
and ending in any species belonging to an apex set. In this section, we shall use the term ``apex predator'' to refer to any member 
of an apex set.

To find the apex sets in a given food web, we make use of {\it random walkers}: imaginary beings that move through the 
network hopping from one node to another along links (in the direction allowed). The walkers are called random because 
at each hop they choose randomly between the different nodes they can access. Random walkers are often used to study 
diffusion processes, and here they can be thought of as representing the diffusion of biomass through the food web. 
Given a network, we simulate many such random walkers beginning at basal species, and for each node we keep a 
register of how many times it has been visited. When the walkers reach an apex set, they cannot leave it, and will 
forever continue to hop around among the members of the group (or they might stay at a single species if it is apex
in the original sense, since there is nowhere to hop). Therefore, whereas most nodes will be visited a small number 
of times which is independent of how long we allow each walker to ``live'', the number of 
times the apex predators are visited increases with walker longevity. This provides a simple computational way of 
finding the apex predators which, though stochastic, will always determine the sets exactly.

Once we know the basal and apex species, we can proceed to find all the chains and obtain the mean value of their 
length.
At least this is possible in principle -- in practice, 
the number is often prohibitively large to be calculated exhaustively. We therefore make use again of the random walkers.
We just have to simulate many walkers beginning at basal species and remove them when they arrive at an apex predator, 
counting how many steps it took them to get there. There is, however, a caveat. The chains actually used provide a 
biased sample of all the chains in the food web: a long chain is more likely to be abandoned somewhere along its length 
than a short one. More precisely, the probability that a particular chain, $\mu$, has of being used is inversely 
proportional to
$$
\pi_\mu=\prod_{i\in \mu}k^{out}_i,
$$
the product extending over all the species $i$ in $\mu$ (except the apex predator), and where $k_i^{out}$ is the number 
of predators of species $i$. So to take this bias into account we calculate, for each walker $w$, not only the length of 
the path it uses, $\lambda_w$, but the value $\pi_{\mu(w)}$, where $\mu(w)$ is the path taken by $w$. After doing this 
for $N$ walkers, an estimate of the mean chain length (which will be more accurate the larger $N$), is
$$
\mbox{MCL}\simeq \frac{\sum_{w=1}^{N} \pi_{\mu(w)}^{-1} \lambda_\mu }{\sum_{w=1}^{N} \pi_{\mu(w)}^{-1}}.
$$
We made sure that this stochastic method converges to the right MCL by comparing the values returned with the results 
from exhaustive searches for those networks where this was possible.

\begin{figure}[ht!]
\renewcommand{\figurename}{Figure S}
\begin{center}
\includegraphics[scale=0.5, angle=270]{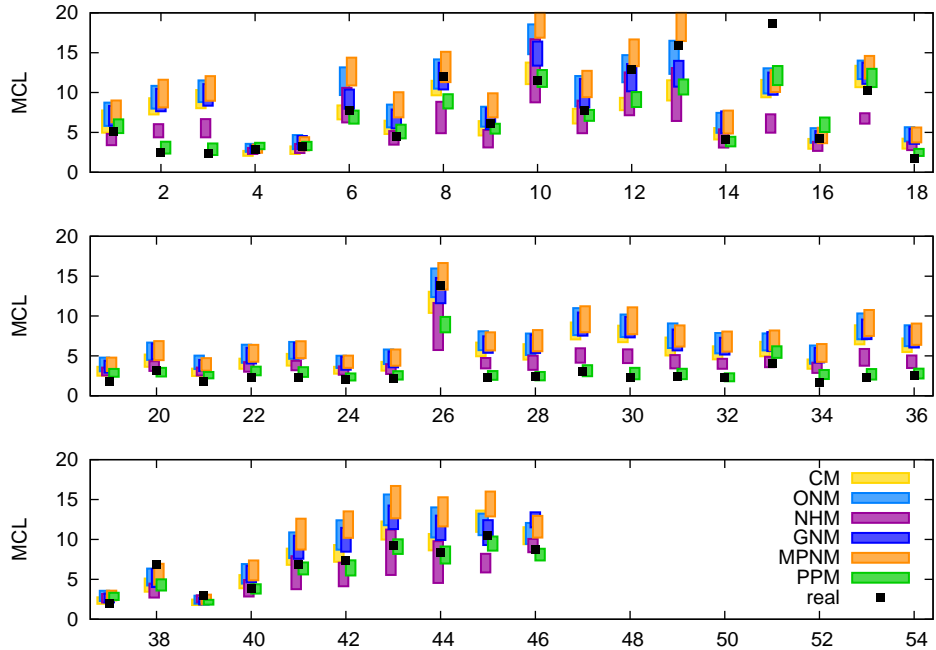}
\end{center}
\caption{
Mean chain length of each of the food webs listed in Table S\ref{table_foodwebs}. The corresponding predictions of each 
food-web model discussed in Section S\ref{Section_Models} -- Cascade, Niche, Nested Hierarchy, Generalized Niche, Minimum Potential Niche
and Preferential Preying -- are displayed with bars representing one standard deviation about the mean. Empirical values are 
black squares. The labelling of the food webs is indicated in the rightmost column of Table S\ref{table_foodwebs}.
}
\label{FigSI_mcl}
\end{figure}

Figure S\ref{FigSI_mcl} shows the predictions of MCL made by each food-web model for the food webs listed in Table S\ref{table_foodwebs}.

\subsection{Modularity}

Much attention has been paid in recent years to the {\it community structure} of complex networks: how the nodes 
can be classified in groups -- or {\it modules} -- such that a high proportion of links fall within groups. 
For a network with $S$ nodes and mean degree $K=L/S$, the {\it configuration model} holds that the 
probability of there being a link from $j$ to $i$ is $k_i^{in} k_j^{out}/(KS)$ 
(where $k_i^{in}$ and $k_i^{out}$ are the numbers of $i$'s prey and predators, respectively)
\cite{Newman_rev}. 
Using this, and given a particular partition (i.e., a classification of nodes into groups) of the network, one can define
$$
Q=\frac{1}{KS}\sum_{ij}\left(A_{ij}-\frac{k_i^{in} k_j^{out}}{KS} \right)\delta(\mu_i,\mu_j),
$$
where $\mu_i$ is a label corresponding to the partition that node $i$ finds itself in, and $\delta(x,y)$ 
is the Kronecker delta \cite{Newman_rev}. The modularity of the network is taken to be the maximum value of $Q$ obtainable 
with any partition. Since searching exhaustively is prohibitive for all but 
very small and sparse networks, a stochastic optimization method is usually called for.
We use the algorithm of Arenas {\it et al.} \cite{Arenas_resolution}, although there are 
many in the literature and the most appropriate can depend on the kind of network at hand \cite{Danon}.

\begin{figure}[ht!]
\renewcommand{\figurename}{Figure S}
\begin{center}
\includegraphics[scale=0.5, angle=270]{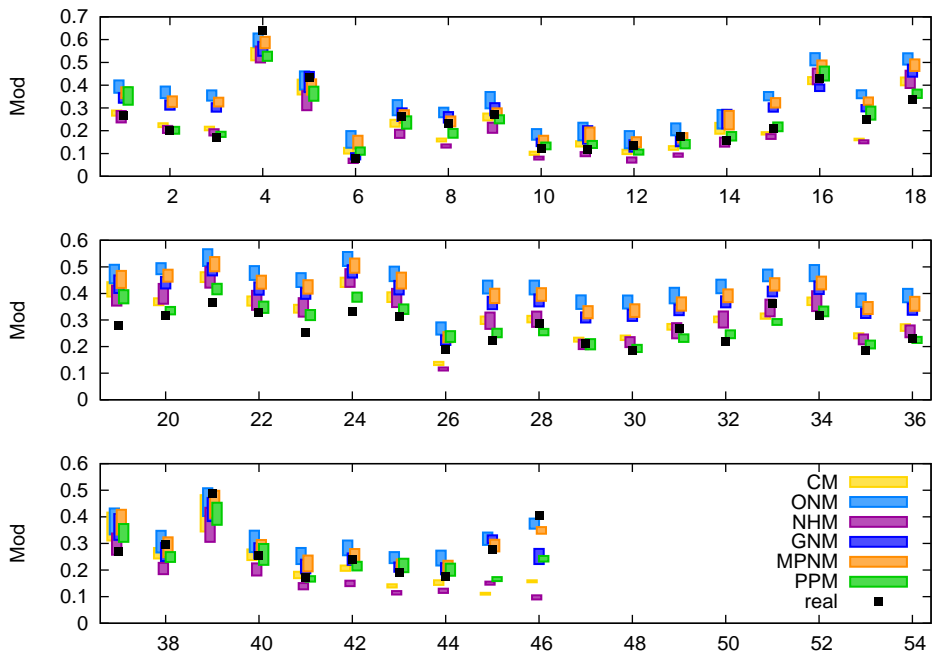}
\end{center}
\caption{
Modularity of each of the food webs listed in Table S\ref{table_foodwebs}. The corresponding predictions of each 
food-web model discussed in Section S\ref{Section_Models} -- Cascade, Niche, Nested Hierarchy, Generalized Niche, Minimum Potential Niche
and Preferential Preying -- are displayed with bars representing one standard deviation about the mean. Empirical values are 
black squares. The labelling of the food webs is indicated in the rightmost column of Table S\ref{table_foodwebs}.
}
\label{FigSI_mod}
\end{figure}

Figure S\ref{FigSI_mod} shows the predictions of modularity made by each food-web model for the food webs listed in 
Table S\ref{table_foodwebs}.

\subsection{Cannibals and apex predators}

\begin{figure}[ht!]
\renewcommand{\figurename}{Figure S}
\begin{center}
\includegraphics[scale=0.5, angle=270]{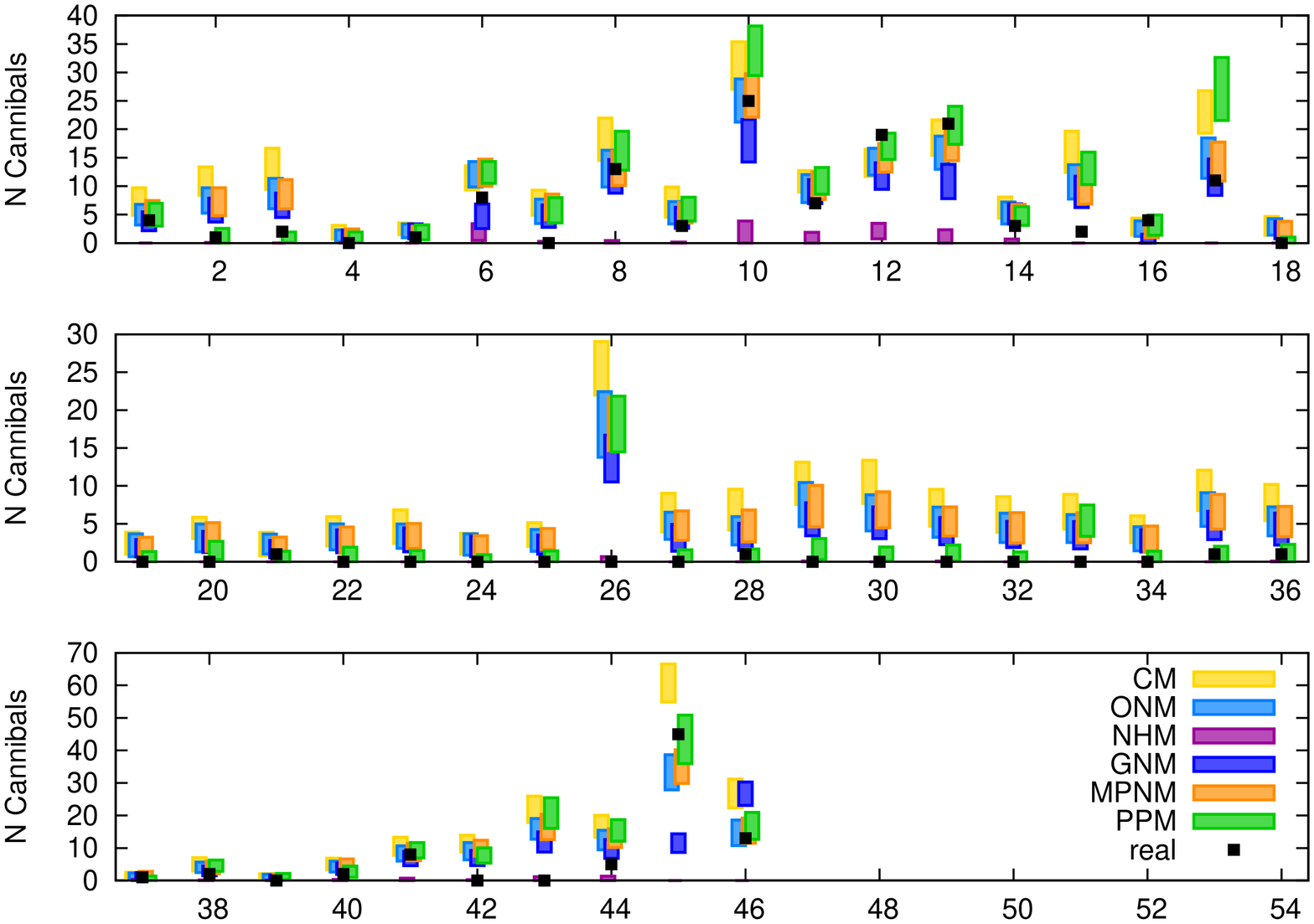}
\end{center}
\caption{
Number of cannibal species in each of the food webs listed in Table S\ref{table_foodwebs}. The corresponding predictions of each 
food-web model discussed in Section S\ref{Section_Models} -- Cascade, Niche, Nested Hierarchy, Generalized Niche, Minimum Potential Niche
and Preferential Preying -- are displayed with bars representing one standard deviation about the mean. Empirical values are 
black squares. The labelling of the food webs is indicated in the rightmost column of Table S\ref{table_foodwebs}.
}
\label{FigSI_canni}
\end{figure}

\begin{figure}[ht!]
\renewcommand{\figurename}{Figure S}
\begin{center}
\includegraphics[scale=0.5, angle=270]{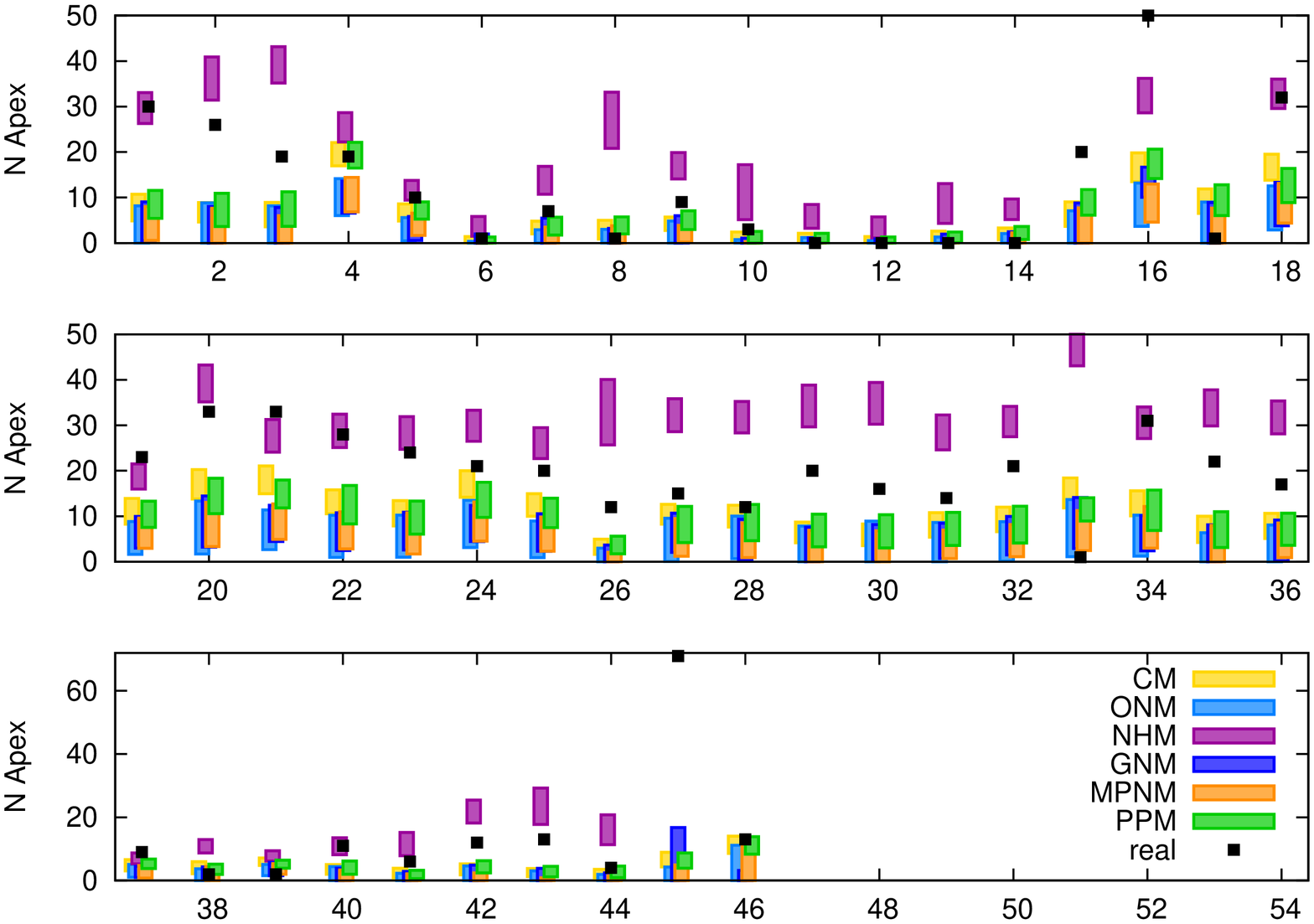}
\end{center}
\caption{
Number of apex predators in each of the food webs listed in Table S\ref{table_foodwebs}. The corresponding predictions of each 
food-web model discussed in Section S\ref{Section_Models} -- Cascade, Niche, Nested Hierarchy, Generalized Niche, Minimum Potential Niche
and Preferential Preying -- are displayed with bars representing one standard deviation about the mean. Empirical values are 
black squares. The labelling of the food webs is indicated in the rightmost column of Table S\ref{table_foodwebs}.
}
\label{FigSI_apex}
\end{figure}

As we have discussed above, cannibalism contributes significantly to stability. We show the number of species with self-links 
predicted by each food-web model for the food webs listed in 
Table S\ref{table_foodwebs} in Fig. S\ref{FigSI_canni}. We also measure the number of apex predators -- in the conventional sense 
of those with no consumers -- and display the model predictions in Fig. S\ref{FigSI_apex}.

\subsection{Mean trophic level}

\begin{figure}[ht!]
\renewcommand{\figurename}{Figure S}
\begin{center}
\includegraphics[scale=0.5, angle=270]{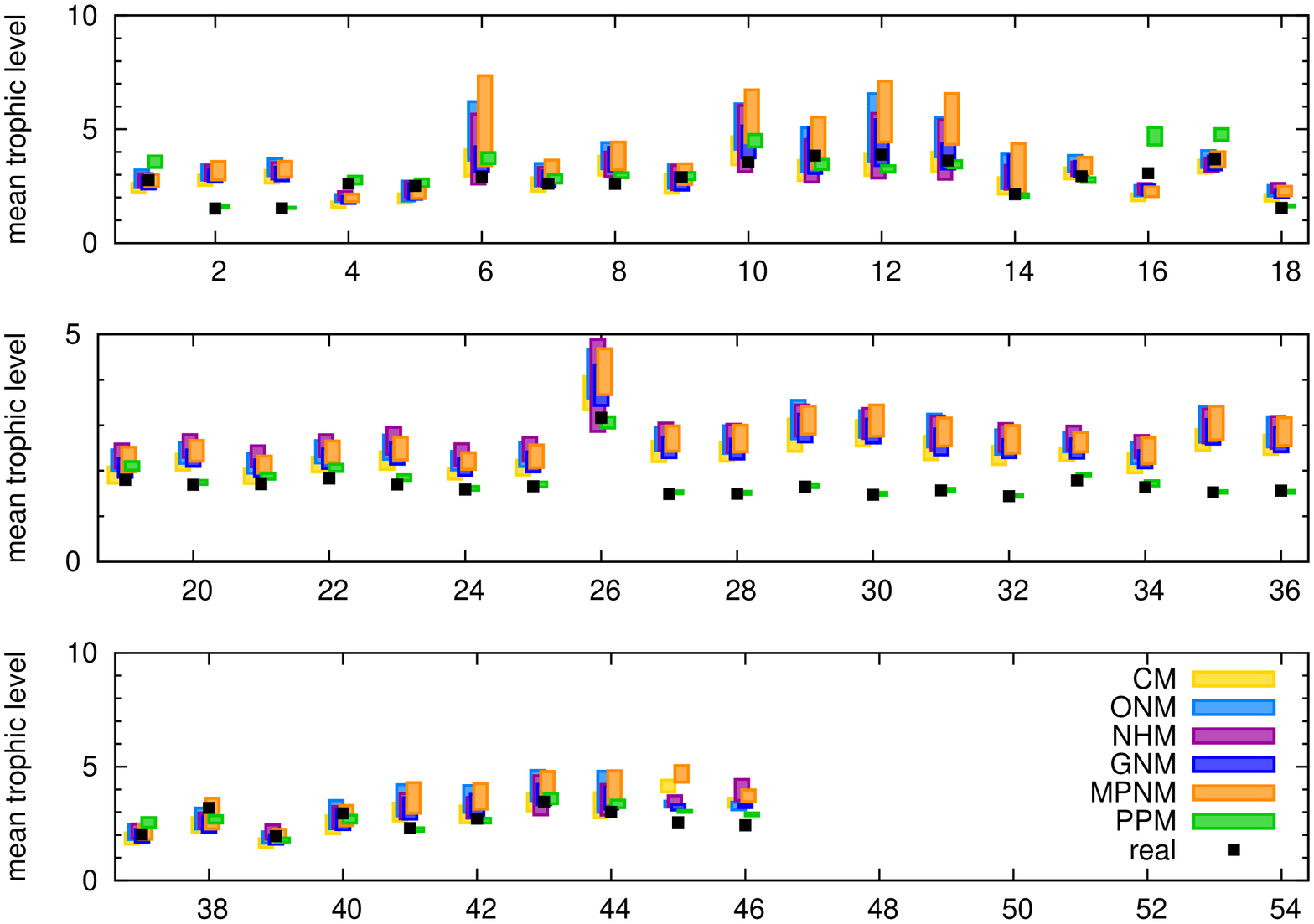}
\end{center}
\caption{
Mean trophic level of each of the food webs listed in Table S\ref{table_foodwebs}. The corresponding predictions of each 
food-web model discussed in Section S\ref{Section_Models} -- Cascade, Niche, Nested Hierarchy, Generalized Niche, Minimum Potential Niche
and Preferential Preying -- are displayed with bars representing one standard deviation about the mean. Empirical values are 
black squares. The labelling of the food webs is indicated in the rightmost column of Table S\ref{table_foodwebs}.
}
\label{FigSI_s}
\end{figure}

The last network feature we analyse is the {\it mean trophic level}, which is simply an average over all the species in a food web of
their trophic levels (i.e., $\overline{s}=S^{-1}\sum_i s_i$). Thanks to Pauly and colleagues' seminal paper ``Fishing down marine food 
webs'' \cite{Pauly_fishing_down}, the mean trophic level has come to be regarded as an indicator of an ecosystem's health, to the extent 
that the Convention on 
Biological Diversity has mandated that signatory states report changes in this measure (renamed the Mean Trophic Index) for
marine ecosystems. The model predictions for the mean trophic level are displayed Fig. S\ref{FigSI_s}.

\subsection{Comparison of network measures}

\begin{figure}[ht!]
\renewcommand{\figurename}{Figure S}
\begin{center}
\includegraphics[scale=0.5, angle=270]{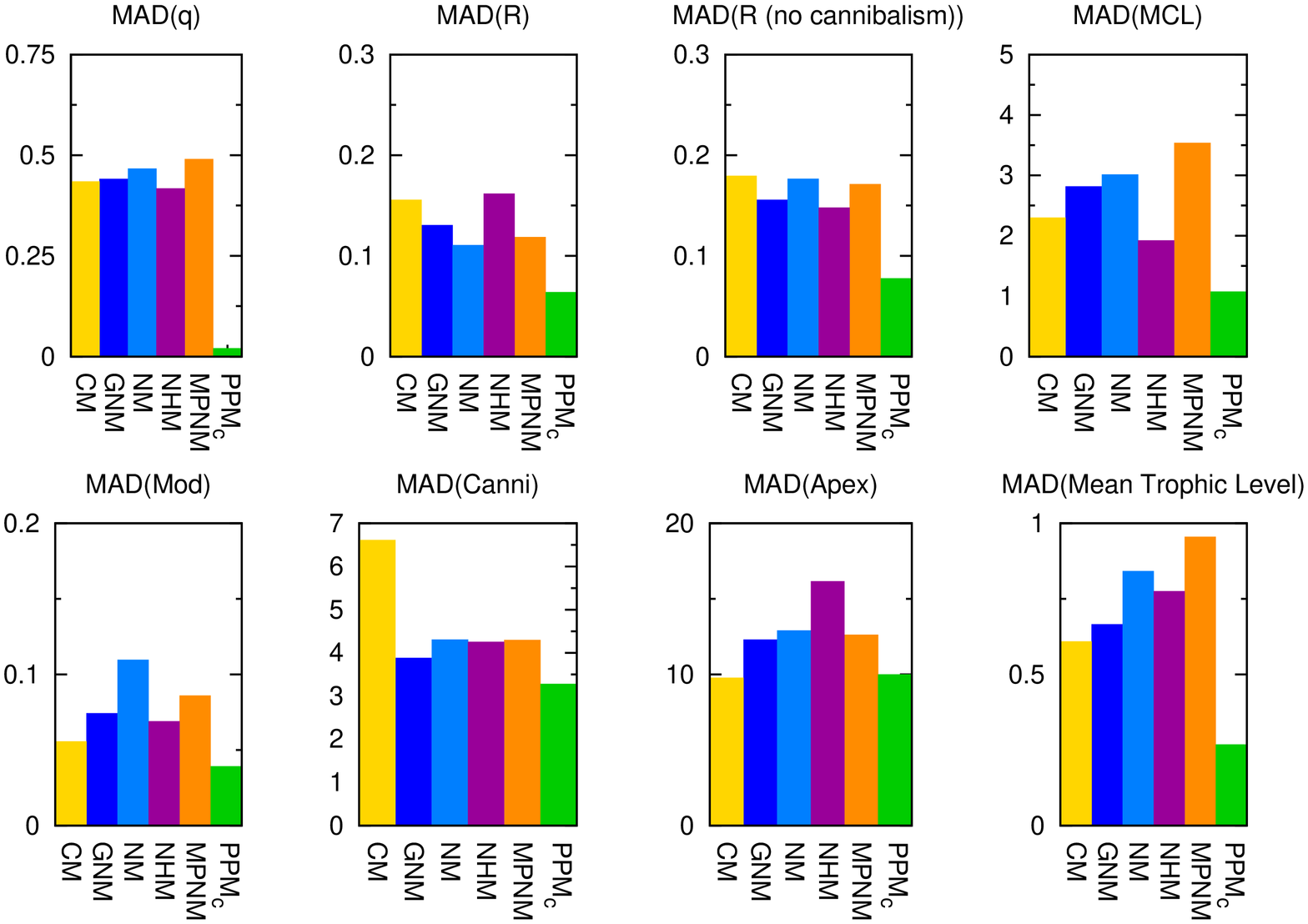}
\end{center}
\caption{
Mean Average Deviation (MAD) form the empirical values returned by each of the food web models discussed in 
Section S\ref{Section_Models} -- Cascade, Niche, Nested Hierarchy, Generalized Niche, Minimum Potential Niche
and Preferential Preying -- for the network measures described in Section S\ref{Section_Models}: trophic coherence $q$, stability $R$,
stability after removing self-links $R_{nc}$, mean chain length, modularity, and numbers of cannibals and of apex predators.
}
\label{FigSI_mad}
\end{figure}

For each of the food-web models and each network measure, we can compute the Mean Average Deviation (MAD) of the theoretical 
prediction, $X_{theo}$ from the empirical value, $X_{empi}$, simply as $MAD=\langle | X_{theo} - X_{empi}| \rangle$, where 
$\langle\cdot\rangle$ stands for an average over the 46 food web listed in Table S\ref{table_foodwebs}. The results for each of the 
eight network measures are shown in the panels of Fig. S\ref{FigSI_mad}. The first panel sums up what we can observe in 
Fig. S\ref{FigSI_q} -- that the niche-based models tend to overestimate the value of $q$ significantly. The fact that none of these 
models differs substantially as regards $q$ from the predictions of the Cascade Model implies that the various features which they 
are designed to capture -- such as intervality, multiple niche dimensions or phylogenetic constraints -- have very little bearing 
on trophic coherence. The Preferential Preying Model, on the other hand, can reproduce the correct value of $q$ in 45 out of 46 food webs
by adjusting its parameter $T$. The odd web out is that of Coachella Valley, which is slightly more incoherent even than the PPM achieves 
with low, negative $T$. This food web is also the only one in our dataset in which more than half the species indulge in cannibalism.
As can be seen from a comparison of Figs. S\ref{FigSI_R} and S\ref{FigSI_Rnc}, this allows the Coachella Valley food web to exhibit 
a relatively low $R$, which it loses when we remove self-links.

The second and third panels show how the models fare as regards stability, both with and without self-links. As discussed in the main 
text, the PPM achieves significantly better results than the other models in both cases, something we attribute to its reproducing
the correct level of trophic coherence. Furthermore, in Figs. S\ref{FigSI_R} and S\ref{FigSI_Rnc} we observe that the niche-based models 
tend to predict less stability than the food webs exhibit, especially in the case without cannibals. This is in keeping with the 
observation by Allesina and Tang \cite{Allesina_Tang} that ``realistic'' food web structure (i.e., that generated with current 
structural models) is not conducive to stability.

Next we look at mean chain length and modularity, two measures which have been associated with ecosystem robustness. In particular, 
a low mean chain length is thought to increase stability \cite{Pimm}, while a high modularity might contain cascades of extinctions
\cite{Stouffer_community}. In keeping with the first observation, the niche-based models tend to predict longer chains than found in 
nature; however, they also somewhat overestimate modularity. In any case, the PPM also outperforms the other models on these two 
measures.

The numbers of cannibals and of apex predators are not very well predicted by any of the models. All but the Nested Hierarchy Model tend 
to overestimate the cannibals and underestimate the apex predators. Finally, we look at the mean trophic level -- a measure which, 
as mentioned above, is used nowadays to assess the health of marine ecosystems and to monitor the effects of 
overfishing \cite{Pauly_fishing_down}. As we might expect from this measure's relationship to trophic structure, the PPM does 
significantly better than the other models at predicting the mean trophic level of food webs. In general, the niche-based models
tend to overestimate the mean trophic level, as shown in Fig. S\ref{FigSI_s}.

\begin{figure}[ht!]
\renewcommand{\figurename}{Figure S}
\begin{center}
\includegraphics[scale=0.5, angle=270]{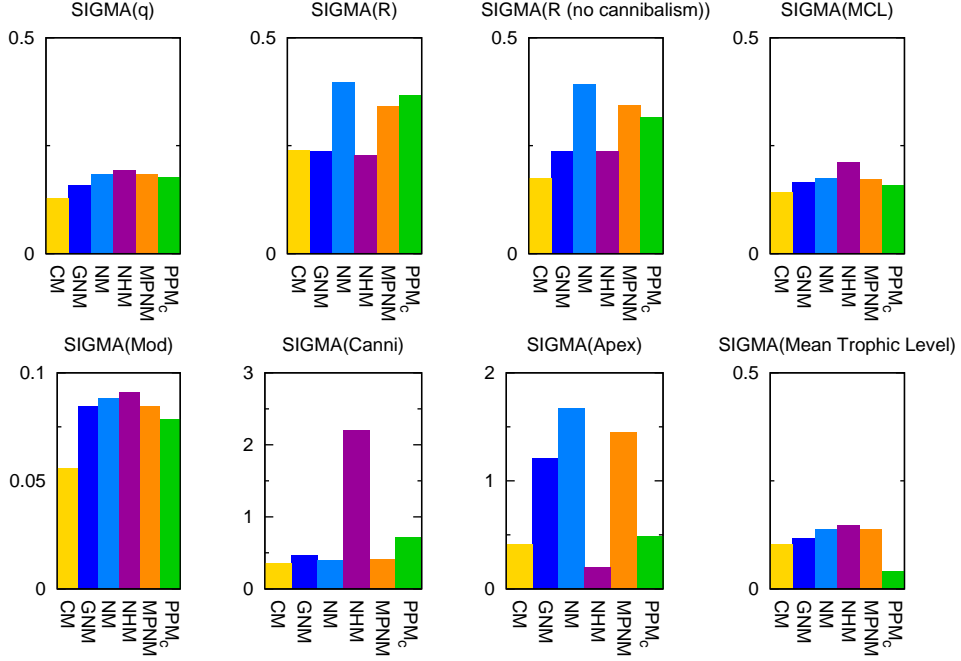}
\end{center}
\caption{
Standard deviation relative to mean, for the 
Mean Average Deviation (MAD) measures displayed in Fig. S\ref{FigSI_mad}.
}
\label{FigSI_sigma}
\end{figure}

\begin{figure}[ht!]
\renewcommand{\figurename}{Figure S}
\begin{center}
\includegraphics[scale=0.5, angle=270]{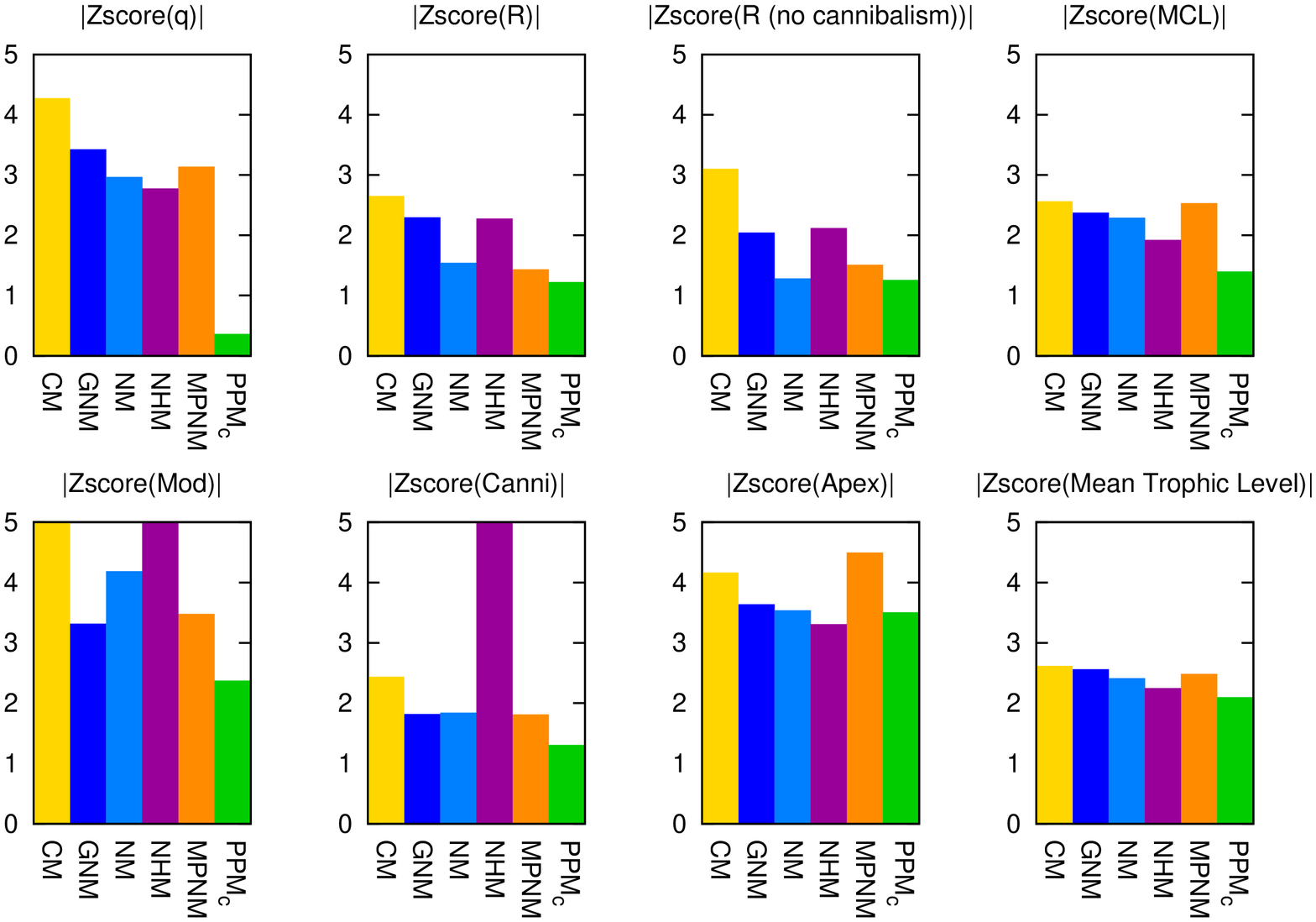}
\end{center}
\caption{
Absolute value of the mean z-score 
returned by each of the food web models discussed in 
Section S\ref{Section_Models} -- Cascade, Niche, Nested Hierarchy, Generalized Niche, Minimum Potential Niche
and Preferential Preying -- for the network measures described in Section S\ref{Section_Models}: trophic coherence $q$, stability $R$,
stability after removing self-links $R_{nc}$, mean chain length, modularity, and numbers of cannibals and of apex predators.
}
\label{FigSI_zscore}
\end{figure}

The standard deviations around the Mean Absolute Deviation measures of Fig. S\ref{FigSI_mad}, relative to each mean value,
are displayed in Fig. S\ref{FigSI_sigma}. 
In Fig. S\ref{FigSI_zscore}, we show the absolute values of the mean z-score obtained for each of the models on the same 
measures.

The comparison we have made here is not as rigorous as one might wish to establish the best food-web model, and this was not our 
intention. For instance, we have not controlled for the number of parameters, nor attempted to 
derive likelihoods for each model, as Allesina {\it et al.} have done \cite{Allesina}. There are also, 
of course, many other network measures of interest in ecology which could be analysed \cite{Capitan,Johnson_nestedness}.
However, we believe it is sufficient to 
show that a) the failure of current structural models to capture trophic coherence is an important shortcoming; and b) 
the Preferential Preying Model, which overcomes this problem, generates networks at least as realistic as any of the other 
structural models. In fact, the PPM significantly outperforms the others on six out of the eight measures we have analysed, and fares 
no worse on the remaining two. However, the PPM does not capture some of the features known to be relevant in food webs, in particular 
a phylogenetic signal \cite{Rossberg_phylogeny}. The high degree of intervality exhibited by many food webs \cite{Stouffer_robust} 
might be a spurious effect of phylogeny and trophic coherence (both of which we know, from preliminary simulations, to contribute 
to intervality) or may need to be modelled explicitly, as in the Niche Model. In any case, we hope to have shown that any attempt 
to build a model which generates networks as similar as possible to real food webs must take account of trophic coherence.

\newpage

\section{Analytical theory for maximally coherent\\networks}
\label{sec_luca}

Let us consider a maximally coherent network, with $q=0$. The $S$ species will thus fall into $M$ discrete 
trophic levels, with $m_i$ species in each level $i$, so that the number of basal species is $B=m_1$, and $S=\sum_{i=1}^M m_i$.
Each link of the predation (or {\it adjacency}) matrix $A$ will 
lead from a prey node at some level $i$ to a predator node a level $i+1$. The interaction matrix
$W=\eta A- A^T$ (where the efficiency $\eta$ is assumed equal for all pairs of species) 
will therefore be an $S\times S$ block matrix where the only nonzero blocks are those 
above and below the main diagonal:
\begin{equation}
  \label{eq:M}
  W =
  \left( \begin{array}{cccccc}
      0 & \eta A_1 & 0 & \ldots & 0 & 0 \\
    -A_1^t & 0 & \eta A_2 & \ldots & 0 & 0 \\
    0 & -A_2^t & 0 & \ldots & 0 & 0 \\
    \hdotsfor{6} \\
    0 & 0 & 0 & \ldots &0 & \eta A_{S-1} \\
    0 & 0 & 0 & \ldots & -A_{S-1}^t & 0 \end{array} \right).
\end{equation}
Blocks $A_i$ are $m_i\times m_{i+1}$ matrices representing the links between the species 
at level $i$ and those at level $i+1$.

Let us now consider the adjacency matrix $\tilde{A}$ of the undirected network we obtain by replacing each directed link 
(or arrow) in $A$ with an undirected (symmetric) one:
\begin{equation}
  \label{eq:A}
  \tilde{A} =
   \left( \begin{array}{cccccc}
    0 &  A_1 & 0 & \ldots & 0 & 0 \\
    A_1^t & 0 & A_2 & \ldots & 0 & 0 \\
    0 & A_2^t & 0 & \ldots & 0 & 0 \\
    \hdotsfor{6} \\
    0 & 0 & 0 & \ldots &0 & A_{S-1} \\
    0 & 0 & 0 & \ldots & A_{S-1}^t & 0
    \end{array} \right).
\end{equation}
The eigenvalues $\{\mu_i\}$ of $\tilde{A}$ are all real since the matrix is symmetric. Furthermore, for every non-negative eigenvalue 
$\mu_j\ge 0$ there is another eigenvalue $\mu_l=-\mu_j$ since the network is bipartite (species can be partitioned into 
two groups with no links within each of them: species in even trophic levels and species in odd levels). Therefore, the 
eigenvalues of $\tilde{A}^2$ are either positive and doubly degenerate or zero.
Moreover, the matrix $\tilde{A}^2$ can be written as:
\begin{equation}
  \label{eq:A2}
  \tilde{A}^2 =
    \left( \begin{array}{ccccc}
    D_1 & 0 & B_1 & 0 & \ldots \\
    0 & D_2 & 0 & B_2 & \ldots \\
    B_1^t & 0 & D_3 & 0 & \ldots \\
    0 & B_2^t & 0 & D_4 & \ldots \\
    \hdotsfor{5}
        \end{array} \right).
\end{equation}
where
\begin{equation}
  \label{eq:blA2}
  \begin{aligned}
    D_i & =
    \begin{cases}
      A_1A_1^t & \text{for $i=1$} \\
      A_{i-1}^tA_{i-1}+A_iA_i^t & \text{for $1<i<M$} \\
      A_{M-1}^tA_{M-1} & \text{for $i=M$}, \\
    \end{cases}\\
    B_i & = A_iA_{i+1}. \\
  \end{aligned}
\end{equation}

Now, the square of matrix $W$ reads:
\begin{equation}
\label{eq:M2}
W^2 =
    \left( \begin{array}{ccccc}
  -\eta D_1 & 0 & \eta^2 B_1 & 0 & \ldots \\
  0 & -\eta D_2 & 0 & \eta^2 B_2 & \ldots \\
  B_1^t & 0 & -\eta D_3 & 0 & \ldots \\
  0 & B_2^t &0 & -\eta D_4 & \ldots \\
  \hdotsfor{5}
  \end{array} \right).
\end{equation}
We introduce a diagonal matrix $U$ with diagonal blocks
\begin{equation}
\label{eq:Ui}
U_{ii}=(-\eta)^{\left\lfloor \frac{i-1}{2}\right\rfloor}I_i,
\end{equation}
where $I_i$ is the identity matrix of size $m_i$, and $\lfloor x\rfloor$ denotes the floor function of $x$:
\begin{equation}
  \label{eq:U}
  U =
     \left( \begin{array}{cccccc}
    I_1 & 0 & 0 & 0 & 0 & \ldots \\
    0 & I_2 & 0 & 0 & 0 & \ldots \\
    0 & 0 & -\eta I_3 & 0 & 0 & \ldots \\
    0 & 0 & 0 & -\eta I_4 & 0 & \ldots \\
    0 & 0 & 0 & 0 & \eta^2 I_5 & \ldots \\
    \hdotsfor{6}
  \end{array} \right).
\end{equation}
We can write
\begin{equation}
  \label{eq:M2A}
   W^2=-\eta U^{-1}\tilde{A}^2U.
\end{equation}
Therefore, the eigenvalues of $W^2$ can be obtained by multiplying those of $\tilde{A}^2$ by $-\eta$: they are either negative and 
doubly degenerate or zero. 
Denoting by $\lambda_j$ the eigenvalues of $W$, we can write
\begin{equation}
  \label{eq:lam2}
  \lambda_j^2 = - \eta \mu_j^2.
\end{equation}

This means that for every $\mu_j=0$ we have $\lambda_j=0$, and for every pair of real eigenvalues $\pm\mu_j$ of $\tilde{A}$ 
there is a pair of imaginary eigenvalues $\lambda_j=\pm \mathrm{i} \sqrt{\eta} \mu_j$ of $W$.
In any case, for $\eta>0$, all the eigenvalues of the interaction matrix $W$ have zero real part. 
If $\eta=0$ all its eigenvalues would be zero, while for $\eta<0$, the imaginary parts would 
vanish and all the eigenvalues would be real, all the nonzero ones coming in pairs $\lambda_j=\pm \sqrt{|\eta|} \mu_j$.

\newpage

\clearpage

\bibliography{refs_fwebs}
\bibliographystyle{ieeetr}

\newpage


